\documentclass[%
superscriptaddress,
 amsmath,amssymb,
 aps,
pra,
]{revtex4-2}

\usepackage{graphicx}
\usepackage{dcolumn}
\usepackage{bm}
\usepackage{bbold}
\usepackage{qcircuit}
\usepackage{xcolor}
\usepackage{amsmath}
\usepackage{amsthm}
\usepackage{multirow}
\usepackage{colortbl}
\usepackage{braket}
\usepackage[normalem]{ulem}
\usepackage{etoolbox}

\newcommand{\Op}[2]{\Ket{#1}\Bra{#2}}

\newcommand{\Zflip}[2]{\hat{\mathcal{Z}}_{#1}^{#2}}

\newcommand{\Cre}{\hat{a}^{\dagger}}
\newcommand{\Ann}{\hat{a}}

\usepackage{kotex}
\usepackage{mathtools, old-arrows}
\usepackage{eqparbox} %
\usepackage{xparse}
\NewDocumentCommand{\eqxrightarrow}{o m O{EO}}{\IfNoValueTF{#1}{\xrightarrow{\eqmakebox[#3]{$\scriptstyle #2$}}}%
{\xrightarrow[#1]{\eqmakebox[#3]{$\scriptstyle #2$}}}}

\begin{document}

\preprint{APS/123-QED}


\title{Doubling Qubits in a Trapped-Ion System via Vibrational Dual-Rail Encoding}

\author{Minhyeok Kang}
\affiliation{SKKU Advanced Institute of Nanotechnology (SAINT), Sungkyunkwan University, Suwon 16419, South Korea}

\author{Wentao Chen}
\affiliation{State Key Laboratory of Low Dimensional Quantum Physics, Department of Physics, Tsinghua University, Beijing 100084, China}

\author{Hyukjoon Kwon}
\email{hjkwon@kias.re.kr}
\affiliation{Korea Institute for Advanced Study, Seoul 02455, South Korea}

\author{Kihwan Kim}
\email{kimkihwan@mail.tsinghua.edu.cn}
\affiliation{State Key Laboratory of Low Dimensional Quantum Physics, Department of Physics, Tsinghua University, Beijing 100084, China}
\affiliation{Beijing Academy of Quantum Information Sciences, Beijing 100193, China}
\affiliation{Frontier Science Center for Quantum Information, Beijing 100084, China}

\author{Joonsuk Huh}
\email{joonsukhuh@yonsei.ac.kr}
\affiliation{Department of Chemistry, Yonsei University,  Seoul 03722, South Korea}




\date{\today}

\begin{abstract}
Vibrational modes of trapped ions have traditionally served as quantum buses to mediate internal qubits. However, with recent advances in quantum control, it has become possible to use these vibrational modes directly as quantum computational resources, such as bosonic qubits. Here, we propose a dual-rail encoding scheme in which a dual-rail qubit is encoded by two vibrational modes that share a single phonon. We present the preparation, measurement, and implementation of single- and two-qubit gates, enabling universal quantum computation. The dual-rail qubit system offers scalability and all-to-all connectivity. Moreover, we extend the dual-rail qubit system to a logical internal qubit--dual-rail qubit hybrid system by incorporating internal qubits into the dual-rail qubit system as another type of logical qubit. The hybrid system nearly doubles the number of available logical qubits compared to conventional trapped-ion quantum computers while maintaining all-to-all connectivity. Additionally, we propose a method for implementing multi-qubit controlled gates and discuss potential applications that can leverage the advantages of the hybrid system. Our scheme provides a practical framework for an internal qubit--boson qubit hybrid system.
\end{abstract}

    \maketitle


\section{Introduction}
\label{sec: introduction}

One method of representing a qubit is to use two quantum systems that share a single excitation. This scheme is known as dual-rail encoding, and the corresponding qubit is called a dual-rail qubit. The computational basis states of the dual-rail qubit \(\ket{0}\) and \(\ket{1}\) are defined by the location of the excitation: \(\ket{0}\) corresponds to the excitation being in the first system, and \(\ket{1}\) to the excitation being in the second. These states span a subspace of the composite Hilbert space, known as the logical subspace, in which the total excitation number is conserved.

For example, in a bosonic system, a dual-rail qubit can be encoded using two bosonic modes and a single boson. The presence of the boson in the first or second mode corresponds to the \(\ket{0}\) or \(\ket{1}\) state of the dual-rail qubit, respectively. Specifically:
\begin{align}
    \ket{0} \Rightarrow  \ket{\underline{1}}\ket{\underline{0}}, \quad \ket{1} \Rightarrow \ket{\underline{0}}\ket{\underline{1}},
\end{align}
where \(\ket{\underline{m}}\) denotes the Fock state of each bosonic mode. This encoding conserves the total number of bosons in the system.

The dual-rail encoding scheme for bosonic systems was first proposed in linear optical network systems, which use photon sources, detectors, beamsplitters, and phase shifters~\cite{chuang1995simple,reck1994experimental}. A dual-rail qubit is encoded in the linear optical network using two optical modes and a single photon. While the linear optical network can realize single-qubit gates through beamsplitters and phase shifters, it cannot perform deterministic two-qubit gate operations~\cite{lutkenhaus1999bell,vaidman1999methods} without nonlinear operations. Although approaches based on cross-Kerr nonlinearity~\cite{chuang1995simple} and measurement-based approaches such as the Knill-Laflamme-Milburn (KLM) protocol~\cite{knill2001scheme} have been proposed to implement two-qubit gates, they face significant limitations: cross-Kerr interactions are typically too weak~\cite{kok2002single}, and the KLM protocol only achieves probabilistic two-qubit gates~\cite{kok2007linear,bartolucci2023fusion}.

Instead of the linear optical network, some alternative platforms for implementing the dual-rail encoding with boson system have been recently explored, such as superconducting platforms~\cite{wallraff2004strong} and trapped-ion systems~\cite{leibfried2003quantum}. Alternative platforms offer a crucial advantage: they can deterministically introduce nonlinear operations via controlled interactions between bosonic modes and qubits---transmon qubits (superconducting qubits based on Josephson junctions) in superconducting circuits and internal qubits (two-level internal states of trapped ions) in trapped-ion systems---enabling universal quantum computation.

A notable implementation has been proposed and demonstrated in superconducting platforms, also known as circuit quantum electrodynamics (circuit QED) systems~\cite{teoh2023dual,levine2024demonstrating,koottandavida2024erasure,crane2024hybrid,chou2024superconducting}. In circuit QED, a dual-rail qubit is encoded using two microwave cavities, each providing a cavity mode, and a single photon. Circuit QED architecture employs beamsplitter couplers and exploits cavity-transmon interactions for state preparation, measurement, and quantum gate operations. Moreover, it enables erasure detection through quantum non-demolition (QND) measurements of photon-number errors and supports scalable circuit construction~\cite{teoh2023dual}. However, its connectivity remains limited to nearest neighbors, making connecting all dual-rail qubits directly difficult.

On the other hand, in a trapped-ion system, a dual-rail qubit can be encoded using two vibrational modes and a single phonon. A trapped-ion phononic network offers significant advantages for dual-rail encoding regarding scalability and connectivity. In a single ion trap with $N$ ions, there are $3N$ modes along the axial and two transverse directions, allowing $N$ dual-rail qubit systems using transverse modes. Nowadays, 20-30 ions in a single ion trap have been applied to perform various quantum algorithms~\cite{chen2024benchmarking}. Thus, dozens of dual-rail qubits can be readily obtained without additional effort. Moreover, since each internal qubit couples to all vibrational modes despite varying strengths~\cite{James1998, marquet2003phonon}, the system achieves native all-to-all connectivity beyond nearest-neighbor interactions. Recent studies have established vibrational modes as quantum resources beyond their traditional role as quantum buses~\cite{lo2015spin,toyoda2015hong,um2016phonon,zhang2018noon,shen2018quantum,fluhmann2019encoding,chen2021quantum,nguyen2021quantum,nguyen2021experimental,gan2020hybrid,chen2023scalable,matsos2024robust,matsos2024universal}, but their potential for dual-rail encoding remains unexplored.

\begin{figure}[htb!]
    \centering
    \includegraphics[width=1.0\linewidth]{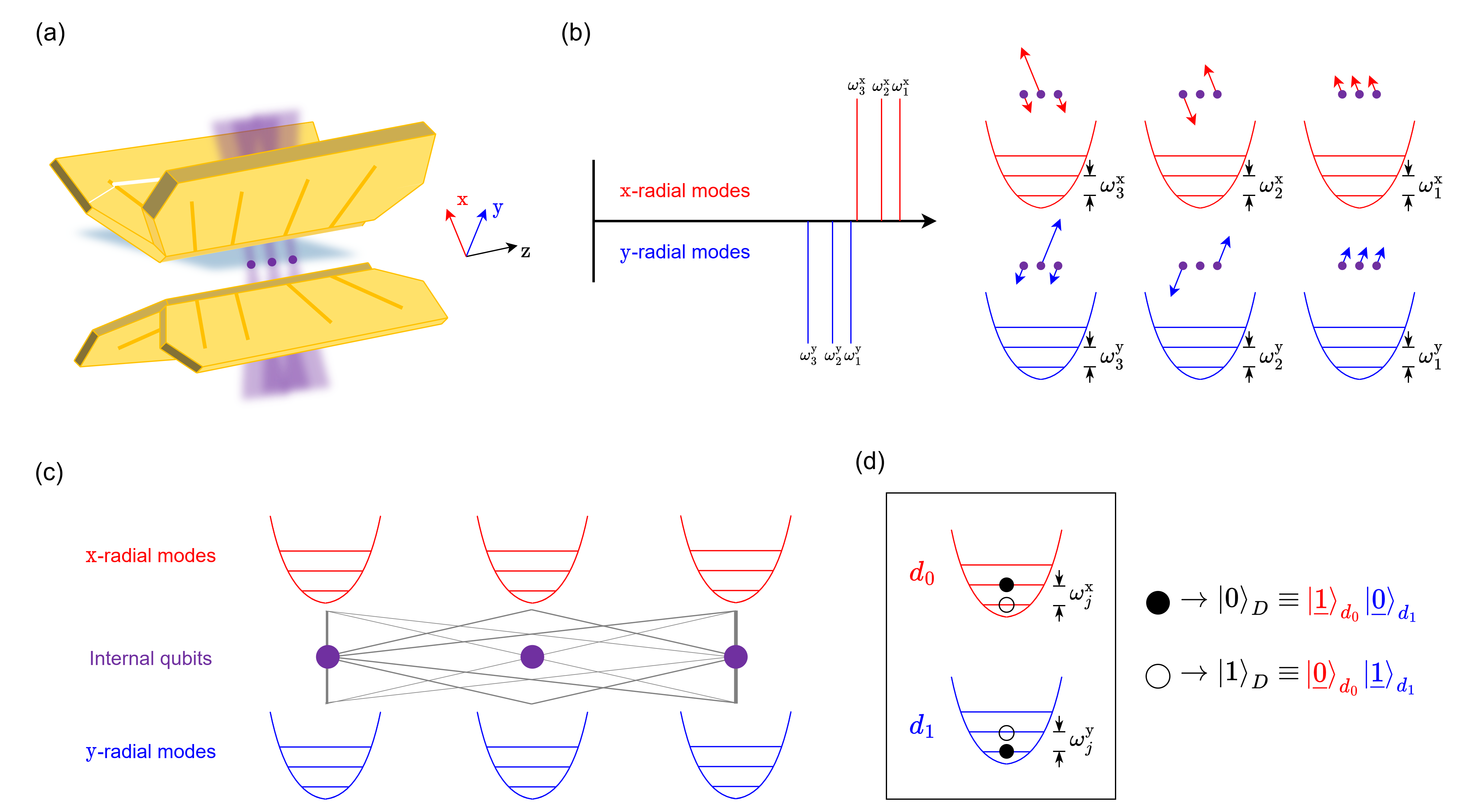}
    \caption{Dual-rail encoding scheme and internal qubit-dual-rail qubit hybrid system using three trapped ions as an example. (a) A typical trapped-ion system for dual-rail encoding featuring individually controllable laser beams for complete manipulation. (b) Illustration of the \(\mathrm{x}\)-radial (red) and \(\mathrm{y}\)-radial (blue) modes of the vibrational spectrum, which are the basis for the dual-rail encoding. On the right, the mode vectors, coupling of individual ions to the modes, are shown for the first three vibration modes, which can be understood as harmonic oscillators with distinct frequencies. (c) Diagram showing the interaction between the internal qubits and transverse modes. The nearly all-to-all connectivity allows internal qubits to interact with multiple vibrational modes, enabling efficient control and gate operations. (d) A dual-rail qubit \(D\) is encoded using two vibrational modes and a single phonon. The computational basis states $\Ket{0}_D$ and $\Ket{1}_D$ of the dual-rail qubit \(D\) are determined by the vibrational mode in which the phonon resides. One example is choosing two modes along different axes.}
    \label{fig: intro_fig}
\end{figure}

Here, we propose the first comprehensive dual-rail encoding scheme using a trapped-ion phononic network. Vibrational modes are controlled through interactions with internal qubits using individually controlled laser beams (Fig.~\ref{fig: intro_fig} (a)). These interactions enable linear and nonlinear operations for state preparation, measurement, and quantum gate implementation on dual-rail qubits ~\cite{leibfried2003quantum, an2015experimental,shen2018quantum}.

In particular, we focus on transverse modes as quantum information carriers for dual-rail encoding. Each transverse mode along a given axis exhibits a collective motion of the ions with a different frequency (Fig.~\ref{fig: intro_fig} (b)). Transverse modes offer several advantages when the number of ions in a single trap increases, since the trap frequencies remain unchanged \cite{kim2009entanglement}. In contrast, forming a linear ion chain with axial modes requires continuously reducing the axial trap frequency, complicating the ground state cooling of axial modes. Moreover, for transverse modes, although the separation between modes decreases as the number of ions increases, they remain manageable even for hundreds of modes \cite{chen2023scalable}.

The strength of the proposed scheme lies in its connectivity architecture. When the ions are tightly confined, nearly all-to-all connectivity is achieved since most internal qubits can interact with the collective vibrational modes~\cite{James1998} (Fig.~\ref{fig: intro_fig} (c)). Consequently, a minimal subset of internal qubits can effectively manipulate all dual-rail qubits, establishing a complete graph topology analogous to internal qubits.

In a trapped-ion phononic network, we encode a dual-rail qubit \(D\) with two transverse modes denoted by \(d_0\) and \(d_1\) with a single phonon. As an example, we can choose two transverse modes along different axes: the mode \(d_0\) from the \(\mathrm{x}\)-radial modes with frequency \(\omega_j^{\mathrm{x}}\) and the mode \(d_1\) from the \(\mathrm{y}\)-radial modes with frequency \(\omega_j^{\mathrm{y}}\), as illustrated in  Fig.~\ref{fig: intro_fig} (d). The computational basis states are then defined by the presence of a single phonon in one of these modes: \(\ket{0}_D = \ket{\underline{1}}_{d_0}\ket{\underline{0}}_{d_1}\) and \(\ket{1}_D = \ket{\underline{0}}_{d_0}\ket{\underline{1}}_{d_1}\).  Although the direction of motion and the frequency differ between the two transverse axes, the overall vibrational spectra and patterns of collective motion remain similar (Fig.~\ref{fig: intro_fig} (b)). As shown in the figure, \(\mathrm{x}\)-radial modes (red) and \(\mathrm{y}\)-radial modes (blue) differ in the direction of motion. Still, both exhibit a collective behavior that resembles each other when considering the overall vibrational structure. Using different axes but the same frequency label \(j\) (or same collective motion) has an advantage in that their coupling strength structures are the same, so we can easily indicate the internal qubit in which they interact.

The implementation of quantum gates for dual-rail qubits utilizes the Pauli Z-dependent beamsplitter (ZBS) gate. The ZBS gate is a hybrid gate acting on an internal qubit and two vibrational modes, applying different beamsplitter operations depending on the basis states of the internal qubit~\cite{shen2018quantum}. The scalability of this gate has been experimentally demonstrated~\cite{chen2023scalable}. A similar approach using the conditional beamsplitter (CBS) gate, which differs from the ZBS gate by only a single beamsplitter operation, has also been demonstrated in a single trapped-ion system~\cite{gan2020hybrid}, highlighting the potential of vibrational modes as dual-rail qubits. However, this demonstration was limited to a single-ion setup, and its scalability remains unexplored. In contrast, our approach naturally extends to multi-ion systems, offering a clear advantage in scalability for large-scale trapped-ion systems.

Since the dual-rail encoding scheme requires a small subset of internal qubits as ancillary resources to manipulate dual-rail qubits, the other internal qubits can function as logical qubits, called logical internal qubits. Thus, the dual-rail encoding scheme can be expanded to form a hybrid system called the logical internal qubit--dual-rail qubit hybrid system, which combines logical internal qubits and dual-rail qubits. Through interactions between internal qubits and vibrational modes, we can implement two-qubit gates between logical internal qubits using a minimal number of ancillary modes and between a logical internal qubit and a dual-rail qubit. Therefore, this hybrid approach enables universal quantum computation by harnessing both types of logical qubits. Moreover, the hybrid system increases the number of available logical qubits in a trapped-ion system without adding ions, while preserving complete graph connectivity.

In this paper, we outline the structure as follows: First, in Sec.~\ref{Sec: dual-rail encoding}, we describe the preparation, measurement, and quantum gate operations for dual-rail qubits. Next, in Sec.~\ref{Sec: hybrid system}, we introduce two-qubit operations between a logical internal qubit and a dual-rail qubit, enabling universal quantum computation in a hybrid system. By leveraging the quantum operations in the hybrid system, we propose an efficient implementation of multi-qubit controlled gates and several applications that benefit from the hybrid system. Finally, in Sec.~\ref{Sec: conclusion}, we present concluding remarks and outline potential directions for future research.

Before proceeding, we define the notations and conventions used throughout this paper to eliminate ambiguity. A summary is provided in Table~\ref{tab: symbol_table}. The internal qubit, formed by two electronic energy levels of an ion, is labeled as \(q\). Its basis states are \(\ket{\downarrow}\) (the ground state) and \(\ket{\uparrow}\) (the excited state). Operators acting on the internal qubit are represented by \(\hat{\sigma}\), including the Pauli operators \(\{\hat{I}, \hat{\sigma}_{\mathrm{x}}, \hat{\sigma}_{\mathrm{y}}, \hat{\sigma}_{\mathrm{z}}\}\) and the raising and lowering operators \(\hat{\sigma}_+\) and \(\hat{\sigma}_-\). Single-qubit rotations on the internal qubit are expressed as \(\hat{\mathcal{R}}\); for example, a rotation about the \(\hat{\sigma}_{\mathrm{x}}\)-axis is written as
\begin{align}
    \hat{\mathcal{R}}_{X}(\theta) = \exp(-i\theta\hat{\sigma}_\mathrm{x}/2).
\end{align}

The bosonic mode (a vibrational mode in a trapped-ion system) is labeled as \(b\). We will simply call it mode. Its Fock states are \(\ket{\underline{m}} = \ket{\underline{0}},\ket{\underline{1}},\cdots\). The annihilation and creation operators are denoted by \(\hat{a}\) and \(\hat{a}^{\dagger}\), respectively. The beamsplitter operator \(\hat{B}(\theta,\phi)\), which performs a 'rotation' between two bosonic modes, is given by
\begin{align}
\label{Eqn: Beamsplitter}
    \hat{B}(\theta,\phi) = \exp \left(i\theta \left(\hat{a}_{b_1}^{\dagger}\hat{a}_{b_2}e^{i\phi}+\hat{a}_{b_1}\hat{a}_{b_2}^{\dagger}e^{-i\phi} \right)\right).
\end{align}

The logical qubit is an abstract qubit used in quantum computation, and its label depends on the specific encoding. The computational basis states are \(\ket{0}\) and \(\ket{1}\), and the associated Pauli operators are \(\{\hat{I},\hat{X},\hat{Y},\hat{Z}\}\). We express single-qubit rotation gates as \(\hat{R}\); for example, an \(X\)-rotation gate is written as
\begin{align}
    \hat{R}_X(\theta) = \exp(-i\theta \hat{X}/2).
\end{align}

\begin{table}[htb!]
    \centering
    \includegraphics[width=0.8\linewidth]{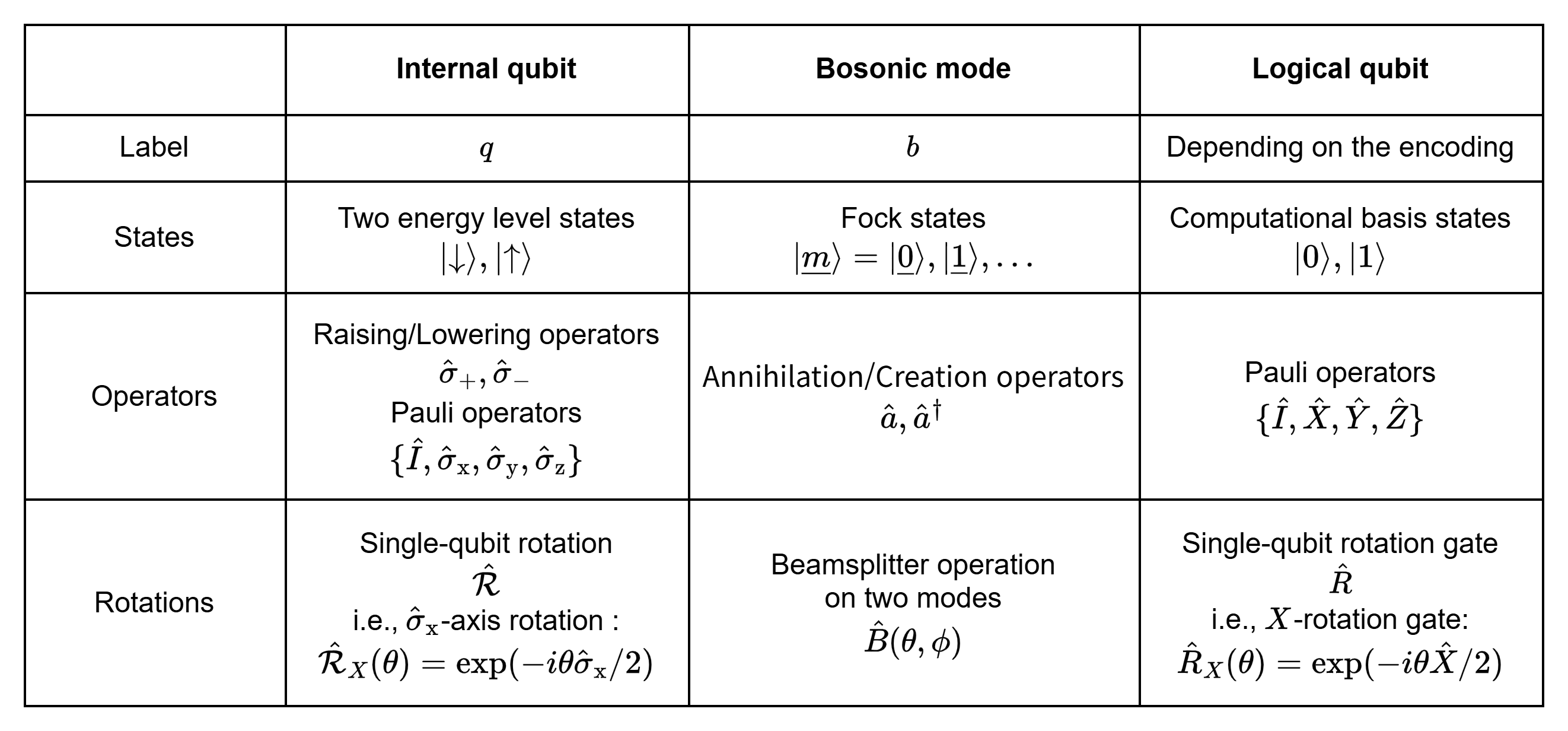}
    \caption{Notations and Conventions for Internal Qubits, Bosonic Modes and Logical Qubits}
    \label{tab: symbol_table}
\end{table}

In our work, the logical qubit is realized in two ways: via the internal qubit and the dual-rail encoding. They are summarized in Table~\ref{tab: encoding_table}. For the internal qubit, the computational basis states are encoded as \(\ket{\downarrow}\Rightarrow \ket{0}\) and \(\ket{\uparrow} \Rightarrow \ket{1}\). We refer to the resulting qubit as the logical internal qubit, and label it by \(q\). The Pauli operators are mapped from the internal qubit operators as follows: \(\hat{\sigma}_{\mathrm{x}} \Rightarrow \hat{X}\),  \(-\hat{\sigma}_{\mathrm{y}} \Rightarrow \hat{Y}\) and  \(-\hat{\sigma}_{\mathrm{z}} \Rightarrow \hat{Z}\). Consequently, \(X\), \(Y\), and \(Z\)-rotation gates for the logical internal qubit are implemented through the following internal qubit rotations:
\begin{gather}
    \hat{\mathcal{R}}_X(\theta) \Rightarrow \hat{R}_X(\theta),\\
    \hat{\mathcal{R}}_Y(-\theta) \Rightarrow \hat{R}_Y(\theta),\\
    \hat{\mathcal{R}}_Z(-\theta) \Rightarrow \hat{R}_Z(\theta).
\end{gather}

A dual-rail qubit is labeled as \(D\), with its constituent bosonic modes denoted by \(d_0\) and \(d_1\). For multiple dual-rail qubits, we use \(D_j\), with modes \(d_{j,0}\) and \(d_{j,1}\). The computational basis states are encoded as \(\ket{\underline{1}}\ket{\underline{0}} \Rightarrow \ket{0}\) and \(\ket{\underline{0}}\ket{\underline{1}}\Rightarrow\ket{1}\). The operator \(\hat{a}^{\dagger}_{d_x}\hat{a}_{d_y}\) corresponds to \(\ket{x}\bra{y}\) for \(x,y \in \{0,1\}\). Accordingly, the beamsplitter operation \(\hat{B}(\theta,\phi)\) implements a rotation gate about the axis \(X\cos\phi + Y\sin\phi\) (see Appendix.~\ref{Appendix_A}):
\begin{align}
\label{eqn: beamsplitter_encoding}
    \hat{B}(\theta,\phi) \Rightarrow \exp(i\theta(\hat{X}\cos\phi+\hat{Y}\sin\phi)).
\end{align}

\begin{table}[htb!]
    \centering
    \includegraphics[width=0.7\linewidth]{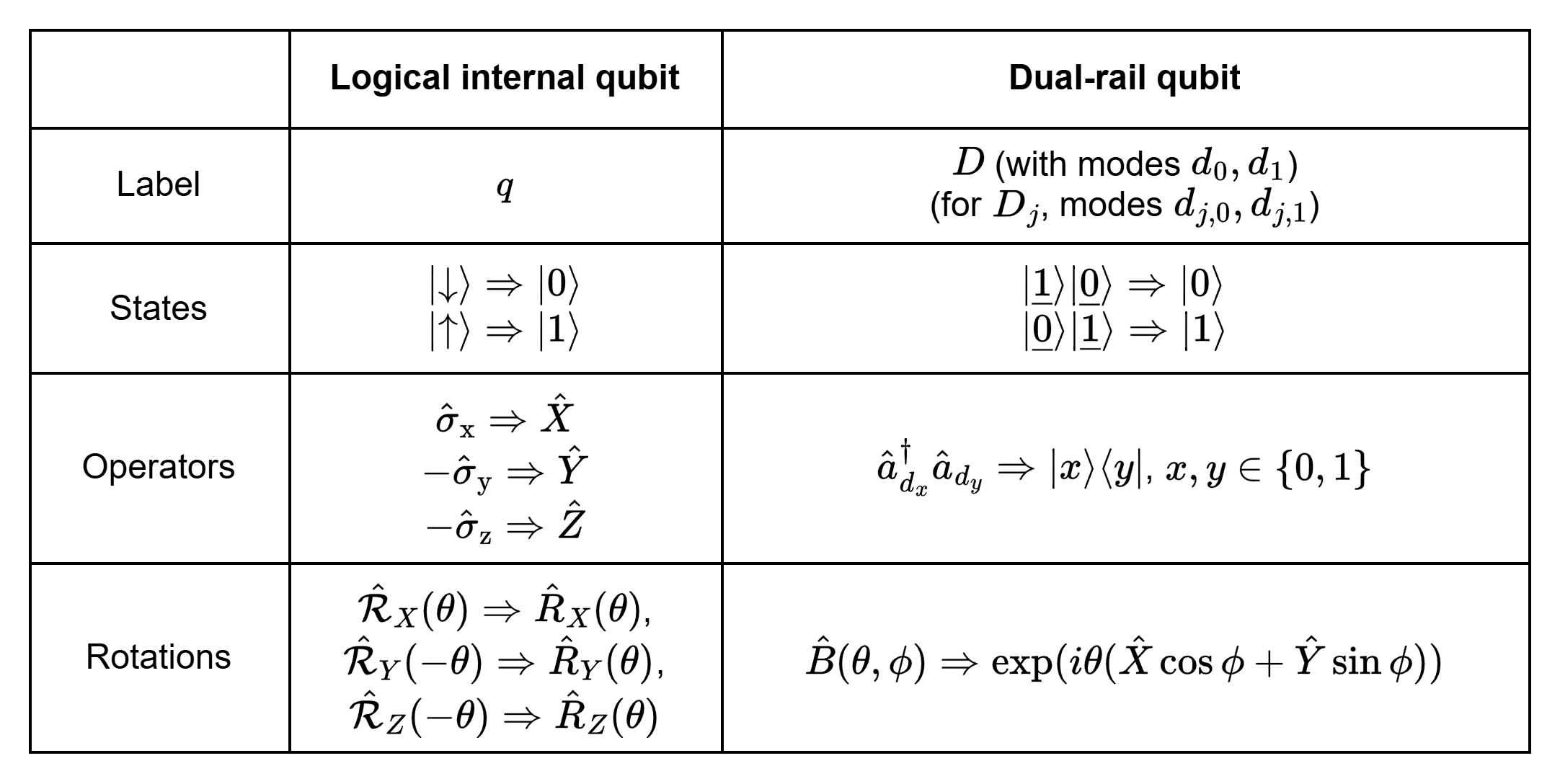}
    \caption{Logical Qubit Encodings and Operator Mappings from Internal Qubits and Bosonic Modes}
    \label{tab: encoding_table}
\end{table}

\section{Dual-rail encoding scheme with trapped-ion phononic network}
\label{Sec: dual-rail encoding}

This section introduces the dual-rail encoding scheme using the vibrational modes in a trapped-ion system to enable universal quantum computation. Overall operations---preparation, measurement, and quantum gate operations---utilize existing interactions between modes and internal qubits. We discuss the properties and advantages of our scheme and compare it with other physical platforms, such as circuit QED~\cite{teoh2023dual}.

Sec.~\ref{Subsec: prep and detect} covers the preparation and measurement of dual-rail qubits. Both rely on carrier and red sideband (RSB) transitions, which are thus discussed together. Following this section, we will assume that all internal qubits are initialized to the ground state, providing a consistent framework for dual-rail qubit operations.

Sec.~\ref{Subsec: manipulation} focuses on the implementation of quantum gates for dual-rail qubits using the ZBS gate. Sec.~\ref{subsubsec: Realization of ZBS} describes the realization of the ZBS gate in a trapped-ion system, specifically, in a system of $^{171}$Yb$^{+}$ ions confined in a Paul trap. Sec.~\ref{subsub: ZBS and single-qubit gate} explores properties of the ZBS gate and shows the implementation of single-qubit gates. Sec.~\ref{subsub: TNP phase gate and two-qubit gate} proposes the implementation of a nonlinear operation which can be used for realizing two-qubit gates. A single ancillary internal qubit is used to apply quantum gates for dual-rail qubits. Since it returns to the ground state after each gate realization, it can be reused in subsequent processes.

Finally, Sec.~\ref{subsec: setting and comparison} presents the properties and advantages of the dual-rail encoding scheme within a trapped-ion phononic network. We also compare our scheme with the superconducting platform, highlighting key differences.

\subsection{Preparation and measurement of the dual-rail qubits}
\label{Subsec: prep and detect}

In the trapped-ion phononic network, computational basis states of dual-rail qubits can be prepared and measured by applying carrier and RSB transitions~\cite{leibfried2003quantum}.

\begin{figure}[htb!]
    \centering
    \includegraphics[width=1.0\linewidth]{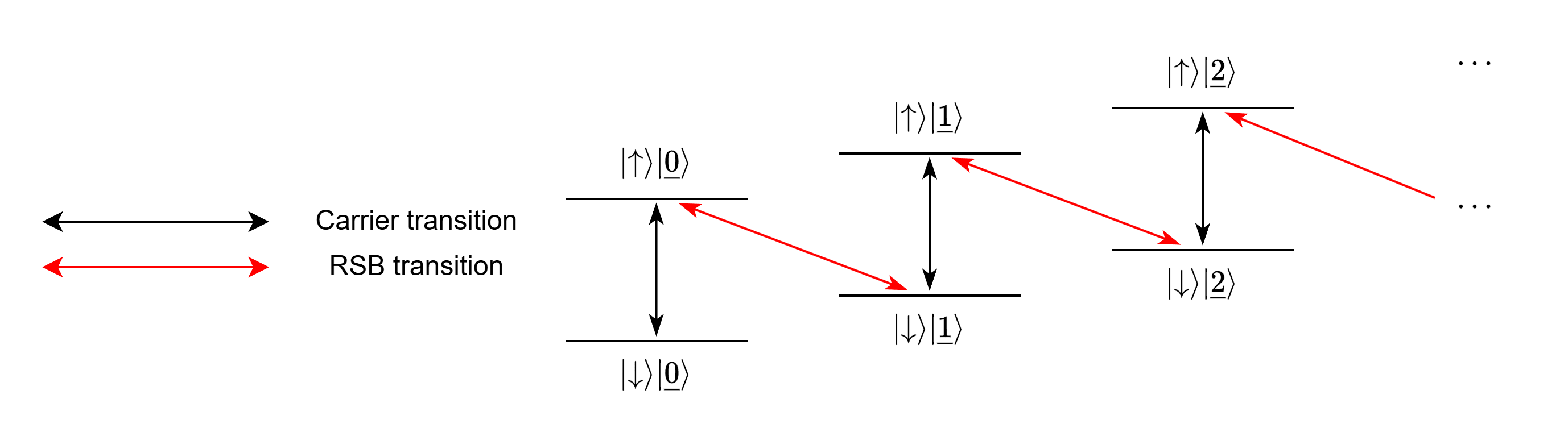}
    \caption{Energy levels of the internal qubit and vibrational mode, illustrating the carrier and RSB transitions. The carrier transition (black) flips the internal qubit state without changing the vibrational mode. The RSB transition (red) couples the states \(\ket{\downarrow}\ket{\underline{m+1}}\) and \(\ket{\uparrow}\ket{{\underline{m}}}\), conserving the total number of excitations in the system.}
    \label{fig: transitions}
\end{figure}

In the trapped-ion system, carrier and RSB transitions are given by (Fig.~\ref{fig: transitions}):
\begin{gather}
\label{eqn: transition}
    \hat{U}_{\rm{carrier}}(\theta,\phi) = \exp{\left( -i\frac{\theta}{2}(\hat{\sigma}_{+}e^{i\phi}+\hat{\sigma}_{-}e^{-i\phi}) \right)},\\
    \hat{U}_{\rm{RSB}}(\theta,\phi) = \exp{ \left(-i\frac{\theta}{2}(\hat{\sigma}_{+}\hat{a} e^{-i\phi}+ \hat{\sigma}_{-}\hat{a}^{\dagger}e^{-i\phi}) \right)}.
\end{gather}
In this paper, we simplify the notation for the RSB transition with \(\phi=0\) as \(\hat{U}_{\rm{RSB}}(\theta) = \hat{U}_{\rm{RSB}}(\theta,0)\). Unless explicitly stated otherwise, the RSB transition will always indicate the case where \(\phi=0\).

The carrier transition (black arrows in Fig.~\ref{fig: transitions}) acts as a single-qubit rotation on the internal qubit, corresponding to the \(\hat{\sigma}_{\mathrm{x}}\cos{\phi} + \hat{\sigma}_{\mathrm{y}}\sin{\phi}\)-axis rotation. Additionally, \(\hat{\sigma}_{\rm{z}}\)-axis rotation on the internal qubit corresponds to the free evolution.

The RSB transition (red arrows in Fig.~\ref{fig: transitions}) arises from the interaction between the bosonic mode and the internal qubit. Notably, the RSB transition plays a crucial role in the dual-rail encoding scheme. It conserves the total number of excitations in the internal qubit and the mode when we interpret \(\hat{\sigma}_{+}\) and \(\hat{a}^{\dagger}\) as the respective raising operators of the two subsystems. Transformations of \(\hat{\sigma}_{+}\) and \(\hat{a}^{\dagger}\)  under the RSB transition are given by:
\begin{gather}
    \hat{U}_{\rm{RSB}}(\theta)\hat{a}^{\dagger}\hat{U}_{\rm{RSB}}^{\dagger}(\theta) = \hat{a}^{\dagger}\cos(\theta/2)- i\hat{\sigma}_{+} \sin(\theta/2),\\
    \hat{U}_{\rm{RSB}}(\theta)\hat{\sigma}_{+}\hat{U}_{\rm{RSB}}^{\dagger}(\theta) = \hat{\sigma}_{+}\cos(\theta/2)-i \hat{a}^{\dagger}\sin(\theta/2).
\end{gather}

For the dual-rail encoding scheme, dual-rail qubits and internal qubits must first be initialized in the \(\ket{0}\) and \(\Ket{\downarrow}\) states, respectively. In particular, initializing the internal qubits in the \(\ket{\downarrow}\) state is essential because we will implement quantum gates for dual-rail qubits using internal qubits prepared in the \(\ket{\downarrow}\) state. As we will show later, the internal qubit involved in the construction of the quantum gate remains in the ground state.

\begin{figure}[htb!]
    \centering
    \includegraphics[width=1.0\linewidth]{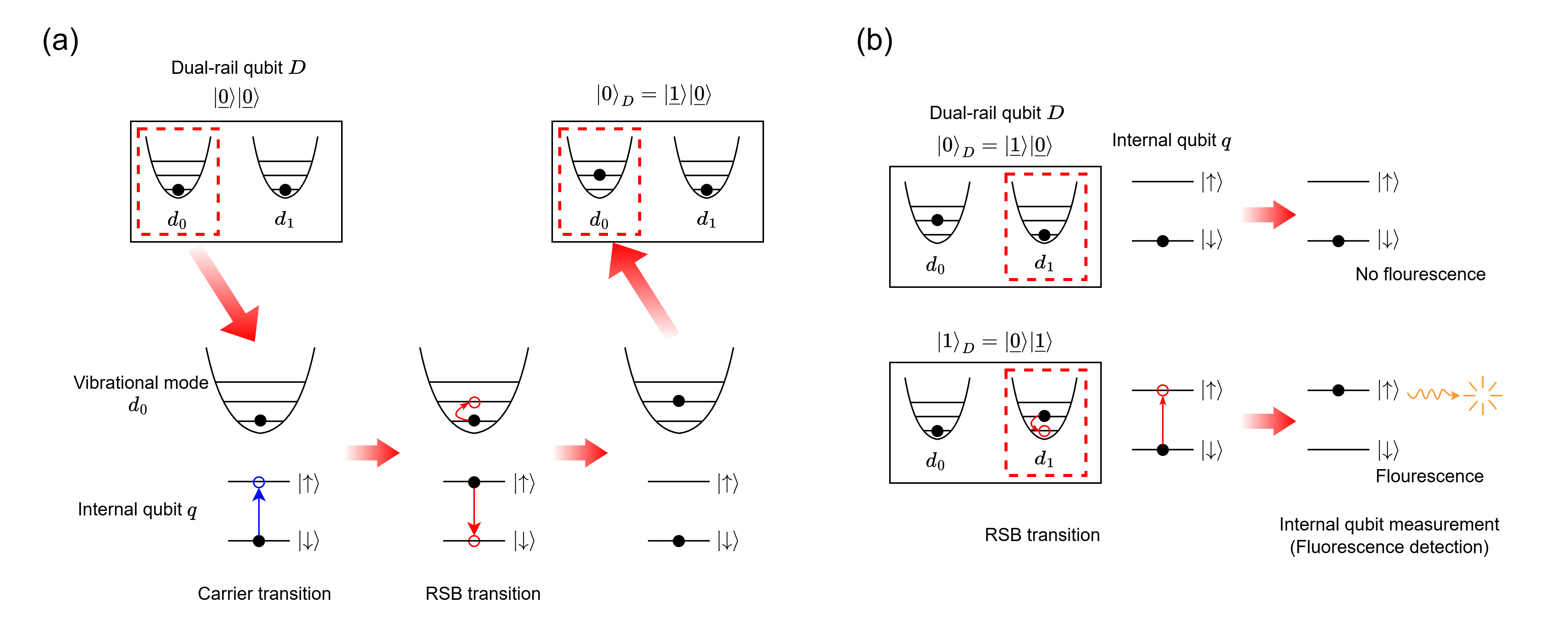}
    \caption{Preparation and measurement of a dual-rail qubit in a trapped-ion system. The dual-rail qubit \(D\) consists of two vibrational modes \((d_0,d_1)\), while \(q\) represents the internal qubit. Both processes assume that \(q\) is initially prepared in the \(\Ket{\downarrow}\) state. (a) Preparation: In the ground state of the trapped-ion system, all modes are in the vacuum state. To initialize the dual-rail qubit \(D\) in the state \(\Ket{0}_D=\ket{\underline{1}}\ket{\underline{0}}\), a single phonon is loaded into mode \(d_0\) via a combination of carrier and RSB transitions. As a result, \(D\) is prepared in \(\Ket{0}_D\), while the internal qubit remains in the \(\Ket{\downarrow}\) state. (b) Measurement: An RSB transition is applied between mode \(d_1\) and the internal qubit \(q\), mapping the quantum state of \(D\) onto \(q\). The dual-rail qubit is then measured by detecting the internal qubit state via fluorescence detection.}
    \label{fig:prep_and_meas}
\end{figure}

Initialization of the trapped-ion system begins by preparing it in its ground state, typically achieved by laser cooling techniques such as Doppler cooling, resolved sideband cooling, and optical pumping operations~\cite{leibfried2003quantum}. After cooling, all internal qubits are in the \(\ket{\downarrow}\) state, and all modes are in the vacuum state. Under this initialization, a dual-rail qubit \(D\) can be prepared in the state \(\ket{0}_D\) by loading a single phonon into mode \(d_0\) using carrier and RSB transitions (Fig.~\ref{fig:prep_and_meas} (a))~\cite{zhang2018noon}.

The measurement of a dual-rail qubit on a computational basis can also be performed using an RSB transition. Consider a dual-rail qubit \(D\) and an internal qubit \(q\) prepared in the \(\ket{\downarrow}\) state. The computational basis state of the dual-rail qubit \(D\) is determined by the presence of a phonon in mode \(d_1\). By applying RSB transition \(\hat{U}_{\mathrm{RSB}}(\pi)\) between mode \(d_1\) and internal qubit \(q\), the quantum state of \(D\) is mapped onto \(q\): \(\ket{0}_D \rightarrow \ket{\downarrow}_q\) and \(\ket{1}_D \rightarrow -i\ket{\uparrow}_q\). Although RSB transition introduces a relative phase in the \(\ket{\uparrow}\) state, it does not affect measurement outcomes in the computational basis. Finally, the measurement is completed via fluorescence detection of the internal qubit~\cite{an2015experimental,um2016phonon,lv2017reconstruction,chen2021quantum}. The measurement outcome of \(q\) directly corresponds to the state of \(D\) (Fig.~\ref{fig:prep_and_meas} (b)).

\subsection{Quantum gate operations for the dual-rail qubits}
\label{Subsec: manipulation}

Universal quantum computation typically requires single- and two-qubit gate operations~\cite{deutsch1995universality}. For dual-rail qubits in a bosonic system, single-qubit gates are implemented using beamsplitter operations, whereas two-qubit gates require both beamsplitter and nonlinear operations.

In phononic networks, quantum gates for dual-rail qubits can be implemented using the hybrid gates induced by interactions between internal qubits and modes. In particular, the ZBS gate plays a central role in constructing quantum gates, since it enables both beamsplitter operations and nonlinear operations acting on modes. Sec.~\ref{subsubsec: Realization of ZBS} discusses the experimental realization of the ZBS gate in a specific trapped-ion system. Sec.~\ref{subsub: ZBS and single-qubit gate} describes the implementation of single-qubit gates on dual-rail qubits using the ZBS gate. Sec.~\ref{subsub: TNP phase gate and two-qubit gate} presents a circuit that combines ZBS gates and single-qubit rotations on internal qubits to perform a nonlinear operation acting on modes, which is the basis of two-qubit gates between dual-rail qubits. As a result, all quantum gate operations for dual-rail qubits can be constructed using ZBS gates together with single-qubit rotations on internal qubits.

\subsubsection{Realization of ZBS in a trapped-ion system}
\label{subsubsec: Realization of ZBS}

The ZBS gate can be implemented in a phononic network using Raman transitions that induce interactions between internal qubits and vibrational modes. In this section, we describe its realization in a system of $^{171}$Yb$^{+}$ ions confined in a Paul trap.

In this system, the two hyperfine clock states in $^{2}$S$_{1/2}$ manifold encode as internal qubit: $\left|\downarrow \right \rangle \equiv |F=0,m_{F}=0 \rangle$ and $\left|\uparrow \right\rangle \equiv |F=1,m_{F}=0 \rangle$ with an energy splitting of $\omega_{q}=12.642812$ GHz. The collective modes are vibrations of these ions, coupled through Coulomb interactions. Collective modes offer significant advantages, such as all-to-all connectivity and low heating rates, making them well-suited for constructing the dual-rail encoding scheme with the trapped-ion phononic network.

\begin{figure}[htb!]
\includegraphics[width=1\textwidth]{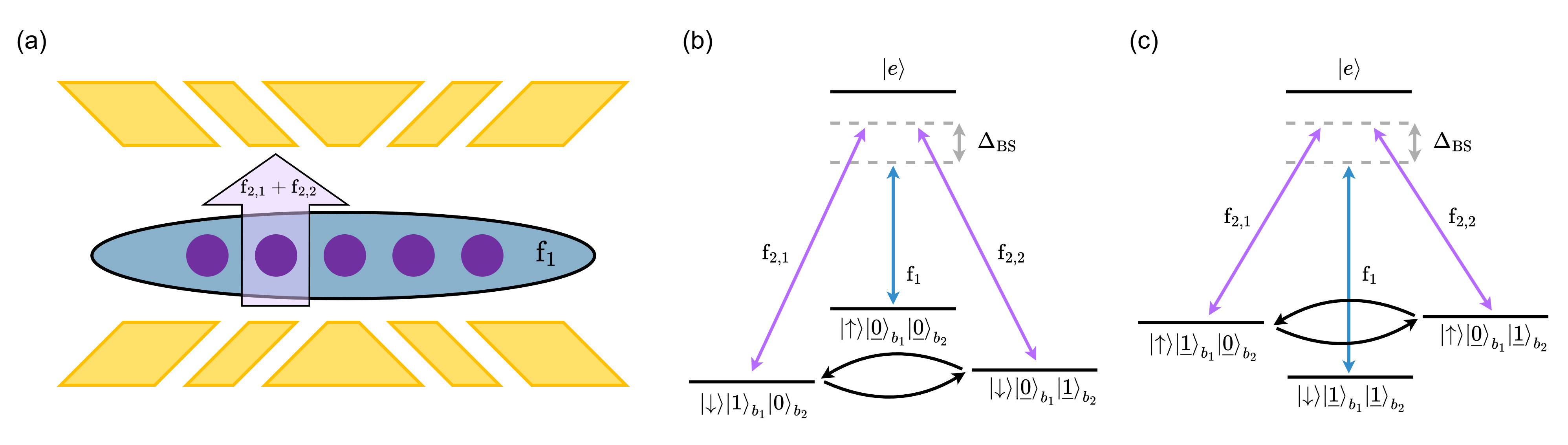}
\caption{Realization of the beamsplitter in a trapped ion system. (a) Raman laser scheme. For a linear chain with five $^{171}$Yb$^{+}$ ions in a blade trap, two Raman lasers are applied simultaneously to realize the beamsplitter. The global laser covers all the ions, and the individual beam focuses on a single ion. (b--c) Energy level scheme for ZBS with qubit energy level in spin-up (down) state.}
\label{Fig: trappedionBS}
\end{figure}

For the realization of the ZBS gate, two pairs of frequencies are simultaneously applied to a single ion, with differences of
\begin{align}
\label{Eqn: CBS frequency}
&\delta_{1} = f_{2,1} - f_1 = \Delta_{\rm{bs}} + \omega_{q} -\nu_{b_1}, \\
&\delta_{2} = f_{2,2} - f_1 = \Delta_{\rm{bs}} + \omega_{q} -\nu_{b_2},
\end{align}
where $\delta_{j}$ denotes the frequency difference for an off-resonant RSB operation of the vibrational mode \(j\). As shown in Fig.~\ref{Fig: trappedionBS} (a), $f_1$ is the frequency of a monochromatic Raman laser, $f_{2,1}$ and $f_{2,2}$ are the frequencies of a bichromatic Raman laser. $\Delta_{\mathrm{bs}}$ denotes the detuning frequency, and $\nu_{b_j}$ denotes the frequency of mode \(j
\). Under the Jaynes-Cummings model, the Hamiltonian of this system takes the form of
\begin{align}
\label{Eqn: CBS Hamiltonian 1}
\hat{H}=\sum_{j=1,2} \frac{\Omega_j}{2}\left( i\eta_{b_j}^{} \hat{\sigma}_{+} \hat{a}_{b_j} e^{-i \Delta_{\rm{bs}} t - i \phi_j} - i\eta_{b_j}^{} \hat{\sigma}_{-} \hat{a}_{b_j}^{\dagger} e^{i \Delta_{\rm{bs}} t + i \phi_j} \right),
\end{align}
where $\Omega_j$ denotes the Rabi frequency of the RSB of vibrational mode \(j\), $\eta_{b_j}$ is the Lamb-Dicke parameter between the chosen ion and mode \(j\), $\phi_j$ denotes the phase of RSB transition on mode \(j\). By ignoring all the oscillating terms in the result, we can get the following effective Hamiltonian
\begin{align}
\label{Eqn: CBS Hamiltonian 2}
\hat{H}_{\rm{ZBS}}=\frac{\eta_{b_1} \eta_{b_2} \Omega_1 \Omega_2}{4 \Delta_{\rm bs}}\hat{\sigma}_{\mathrm{z}}(\hat{a}_{b_1}^{\dagger} \hat{a}_{b_2}^{} e^{i(\phi_1-\phi_2)} + \hat{a}_{b_1}^{} \hat{a}_{b_2}^{\dagger} e^{-i(\phi_1-\phi_2)}),
\end{align}
which is used for a standard ZBS operation as shown in Fig. \ref{Fig: trappedionBS} (b) and (c).

\subsubsection{ZBS gate and single-qubit gate for dual-rail qubits}
\label{subsub: ZBS and single-qubit gate}

\begin{figure}[htb!]
    \centering
    \includegraphics[width=1.0\linewidth]{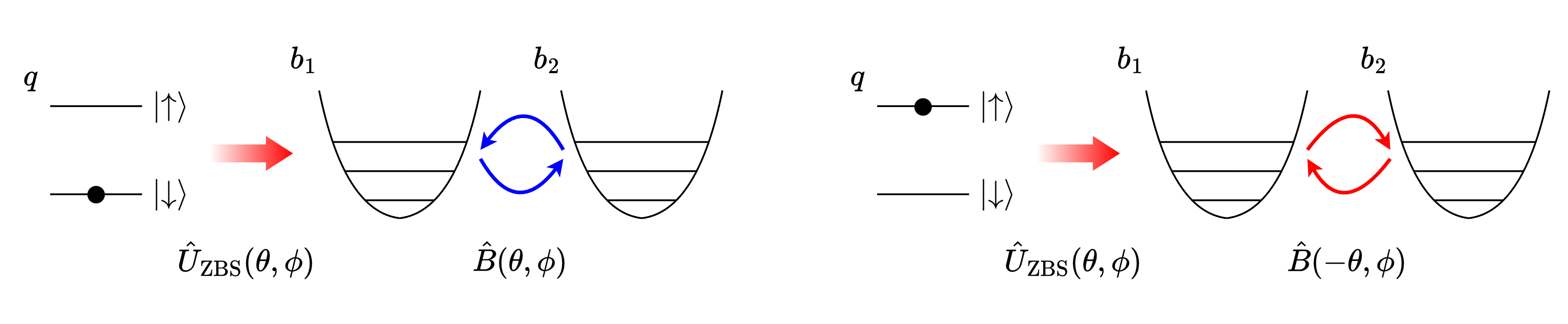}
    \caption{Graphical representation of the ZBS gate. When the ZBS gate is applied to two modes and an internal qubit, the basis state of the internal qubit determines the sign of the parameter \(\theta\) of the beamsplitter operation acting on the modes.}
    \label{fig:ZBS_graphical}
\end{figure}

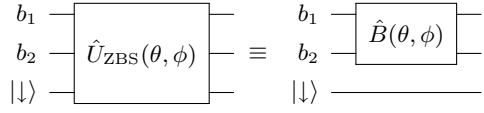
\begin{figure}
    \centering
    \[ \Qcircuit @C=1em @R=.7em{
            \lstick{b_1}&\multigate{2}{\hat{U}_{\text{ZBS}}(\theta,\phi)} & \qw &  &&& \lstick{b_1} & \multigate{1}{\hat{B}(\theta,\phi)} & \qw\\
            \lstick{b_2}& \ghost{\hat{U}_{\text{ZBS}}(\theta,\phi)} & \qw & \equiv &&& \lstick{b_2} & \ghost{\hat{B}(\theta,\phi)} & \qw\\
            \lstick{\Ket{\downarrow}} &\ghost{\hat{U}_{\text{ZBS}}(\theta,\phi)} & \qw  & &&& \lstick{\Ket{\downarrow}}  & \qw & \qw\\
            }
        \]
    \caption{Implementation of a beamsplitter operation using the ZBS gate. When the internal qubit is prepared in a computational basis state, the ZBS gate implements a beamsplitter operation in a phononic network. In this case, the internal qubit is initialized in the ground state \(\Ket{\downarrow}\).}
    \label{Fig: BS by ZBS}
\end{figure}

From Sec.~\ref{subsubsec: Realization of ZBS}, we can express the ZBS gate based on the effective Hamiltonian given in Eq.~\ref{Eqn: CBS Hamiltonian 2}. The ZBS gate is written as:
\begin{gather} \label{Eqn: ZBS}
    \hat{U}_{\text{ZBS}}(\theta,\phi) = \exp{\left(-i\theta\hat{\sigma}_{\mathrm{z}} \left(\hat{a}_{b_1}^{\dagger}\hat{a}_{b_2}e^{i\phi}+\hat{a}_{b_1}\hat{a}_{b_2}^{\dagger}e^{-i\phi}\right)\right)},\\
    \label{Eqn:param}
    \theta := \frac{\eta_{b_1} \eta_{b_2} \Omega_1 \Omega_2}{4 \Delta_{\rm bs}}t, \quad \phi:=\phi_1-\phi_2.
\end{gather}
In Eq.~\ref{Eqn:param}, \(t\) is the propagation time of the ZBS gate. The parameter \(\theta\) can be adjusted by controlling \(t\). Due to the presence of \(\hat{\sigma}_{\mathrm{z}}\), the sign of the parameter \(\theta\) in the beamsplitter operator acting on the modes depends on the basis state of the internal qubit (i.e., its eigenstate with respect to \(\hat{\sigma}_\mathrm{z}\)) when the ZBS gate is applied (Fig.~\ref{fig:ZBS_graphical}):
\begin{gather}
    \label{eqn: ZBS 2}
    \hat{U}_{\text{ZBS}}(\theta,\phi)\Ket{\sigma}\Ket{\psi} = \Ket{\sigma} \hat{B}(-\lambda_\sigma\theta,\phi)\Ket{\psi},\\
\end{gather}
where \(\sigma \in \{\uparrow,\downarrow\}\), \(\lambda_\sigma\) is the eigenvalue of \(\hat{\sigma}_{\mathrm{z}}\), i.e., $\hat{\sigma}_{\mathrm{z}} \ket{\sigma} = \lambda_\sigma \ket{\sigma}$, \(\ket{\psi}\) denotes a quantum state of the two modes, and \(\hat{B}(\theta,\phi)\) is the beamsplitter operator on the modes. Note that \(\lambda_{\downarrow}= -1\) and \(\lambda_{\uparrow}=+1\). Accordingly, the ZBS gate can be rewritten as:
\label{eqn: ZBS_split}
\begin{align}
    \hat{U}_{\mathrm{ZBS}}(\theta,\phi) = \ket{\downarrow}\bra{\downarrow}\otimes\hat{B}(\theta,\phi) + \ket{\uparrow}\bra{\uparrow}\otimes\hat{B}(-\theta,\phi).
\end{align}
A beamsplitter operation acting on the modes can be straightforwardly implemented by applying a ZBS gate to an internal qubit prepared in an eigenstate of \(\hat{\sigma}_\mathrm{z}\)~\cite{chen2023scalable} (Fig.~\ref{Fig: BS by ZBS}). Since the beamsplitter operation implements \(X\cos\phi + Y\sin\phi\)-rotation gate for dual-rail qubit, it can be used to realize an arbitrary single-qubit gate for dual-rail qubits (See Appendix.~\ref{Appendix_A}).

Furthermore, the ZBS gate facilitates nonlinear operations in the phononic network when combined with single-qubit rotations on the internal qubit. This capability allows the construction of two-qubit gates between dual-rail qubits, which will be discussed in detail in the next section.

\subsubsection{Total phonon number parity dependent phase gate and Two-qubit gate for dual-rail qubits}
\label{subsub: TNP phase gate and two-qubit gate}

\begin{figure}[htb!]
\centering
\[ \Qcircuit @C=1em @R=.7em{
    \lstick{\Ket{\underline{n}}}        & \qw &  \multigate{2}{\hat{\mathcal{Z}}} & \qw & \multigate{2}{\hat{\mathcal{Z}}^{-1}} & \qw & \qw \\
    \lstick{\Ket{\underline{m}}}  & \qw & \ghost{\hat{\mathcal{Z}}} & \qw & \ghost{\hat{\mathcal{Z}}^{-1}} & \qw  & \qw & \rstick{\raisebox{1.8em} {$\exp{\left(i\frac{(-1)^{n+m+1}\theta}{2}\right)}\Ket{\underline{n}}\Ket{\underline{m}}  $} }\\
    \lstick{\Ket{\downarrow}}   & \gate{\hat{\mathcal{R}}_Y(-\pi/2)} & \ghost{\hat{\mathcal{Z}}} & \gate{\hat{\mathcal{R}}_X(\theta)} & \ghost{\hat{\mathcal{Z}}^{-1}} & \gate{\hat{\mathcal{R}}_Y(\pi/2)}  & \qw & \Ket{\downarrow} \\
    {\gategroup{1}{7}{2}{7}{1.em}{\}} }
} \]
\caption{TNP phase gate using ZBS and single-qubit gates on the internal qubit. The sign of the resulting phase angle depends on the total phonon-number parity of the Fock states in the two modes. When applied to the vibrational modes of each dual-rail qubit, this circuit implements a \(ZZ\)-rotation gate on dual-rail qubits.}
\label{Fig: total number parity phase gate}
\end{figure}
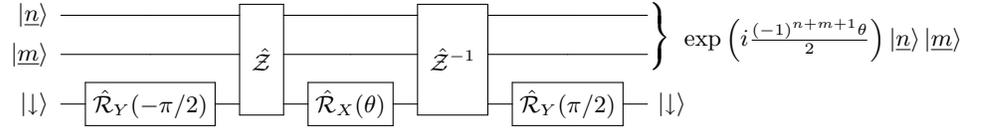

ZBS gates and single-qubit rotations on the internal qubit can be used to implement two-qubit gates for dual-rail qubits. Consider the effect of a specific ZBS operation, defined as \(\hat{\mathcal{Z}}:= \hat{U}_{\text{ZBS}}(\pi/2,0)\), acting on a product state consisting of the basis state of the internal qubit and the Fock states of the two modes:
\begin{align}
\label{Eqn: Zflip properties}
    \hat{\mathcal{Z}}\Ket{\sigma}\Ket{\underline{n}}\Ket{\underline{m}} = (-\lambda_\sigma i)^{n+m}\Ket{\sigma}\Ket{\underline{m}}\Ket{\underline{n}} = (-\lambda_\sigma)^{(n+m)}i^{n+m}\Ket{\sigma}\Ket{\underline{m}}\Ket{\underline{n}},
\end{align}
where \(\sigma \in \{\downarrow, \uparrow\}\). The operation \(\hat{\mathcal{Z}}\) swaps the quantum states of the two modes with a phase factor \((-\lambda_\sigma)^{(n+m)}i^{n+m}\). The term \((-\lambda_\sigma)^{(n+m)}\) encodes the total phonon number parity of the two modes in the internal qubit. When the internal qubit is initialized in the state \(\frac{1}{\sqrt{2}}(\Ket{\downarrow} + \Ket{\uparrow})\), the \(\Zflip{}{}\) gate swaps the states of the two modes with a phase factor of \(i^{n+m}\), and the internal qubit state transforms into \(\frac{1}{\sqrt{2}}(\ket{\downarrow}+\ket{\uparrow})\) or \(\frac{1}{\sqrt{2}}(\ket{\downarrow}-\ket{\uparrow})\), depending on whether \(n+m\) is even or odd, respectively.
By combining this operation with \(\hat{\sigma}_{\rm{x}}\)-axis and \(\hat{\sigma}_{\mathrm{y}}\)-axis rotations on the internal qubit, the circuit shown in Fig.~\ref{Fig: total number parity phase gate} transforms the initial state \(\Ket{\downarrow}\Ket{\underline{n}}\Ket{\underline{m}}\):
\begin{align}
\label{Eqn: total number parity phase gate}
    \Ket{\downarrow}\Ket{\underline{n}}\Ket{\underline{m}} \xrightarrow{\hat{\mathcal{R}}_Y(\pi/2) \hat{\mathcal{Z}}^{-1}\hat{\mathcal{R}}_X(\theta)\hat{\mathcal{Z}}\hat{\mathcal{R}}_{Y}(-\pi/2)}
    \exp{\left(i\frac{(-1)^{(n+m+1)}\theta}{2}\right)}\Ket{\downarrow}\Ket{\underline{n}}\Ket{\underline{m}}.
\end{align}
In the resulting state, the internal qubit remains unchanged, while a global phase is applied that depends on the total phonon number parity of the two modes. We refer to this operation as the total number parity-dependent (TNP) phase gate (see Appendix~\ref{Appendix: total number parity phase} for an explicit derivation). 

The TNP phase gate is a \(ZZ\)-rotation (RZZ) gate for dual-rail qubits, \(\hat{R}_{ZZ}(\theta)=e^{-i\theta \hat{Z}\hat{Z}/2}\). Let \(D_1\) and \(D_2\) be dual-rail qubits. Since the number of phonons in mode \(d_{i,1}\) determines the computational basis state of the dual-rail qubit \(D_i\), applying the TNP phase gate on the two modes \((d_{1,1},d_{2,1})\) yields the following transformations:
\begin{align}
    &\Ket{0}_{D_1}\Ket{0}_{D_2} \rightarrow e^{-i\frac{\theta}{2}}\Ket{0}_{D_1}\Ket{0}_{D_2}, &\Ket{1}_{D_1}\Ket{1}_{D_2} \rightarrow e^{-i\frac{\theta}{2}}\Ket{1}_{D_1}\Ket{1}_{D_2},\\
    &\Ket{1}_{D_1}\Ket{0}_{D_2} \rightarrow e^{i\frac{\theta}{2}}\Ket{1}_{D_1}\Ket{0}_{D_2}, &\Ket{0}_{D_1}\Ket{1}_{D_2} \rightarrow e^{i\frac{\theta}{2}}\Ket{0}_{D_1}\Ket{1}_{D_2}.
\end{align}
Moreover, the RZZ gate \(\hat{R}_{ZZ}(\pi/2)\) is equivalent to a Controlled-NOT (CNOT) gate up to single-qubit gates~\cite{makhlin2002nonlocal,schuch2003natural}.

\subsection{Properties of dual-rail qubit system on trapped-ion system and comparison with superconducting platform}
\label{subsec: setting and comparison}

In Secs.~\ref{Subsec: prep and detect} and \ref{Subsec: manipulation}, we have described the preparation, measurement, and quantum gate operations for dual-rail qubits within the phononic network in a trapped-ion system, demonstrating their capability for universal quantum computation. In this section, we examine the properties of this platform and provide a detailed comparison with the superconducting cavity platform.

In a trapped-ion system, there are two types of quantum systems: vibrational modes and internal qubits. A vibrational mode corresponds to the harmonic oscillation of ions and is inherent to this system, while an internal qubit consists of two electronic energy levels within an ion.

As described in Introduction, there are \(3N\) modes along the axial and two transverse directions in a single ion trap containing \(N\) ions. Since the number of available transverse modes scales with the number of trapped ions, increasing the number of ions also increases the number of dual-rail qubits, providing scalability. Furthermore, all operations for the dual-rail encoding scheme can be implemented using existing Raman beam techniques, eliminating the need for additional components.

\begin{figure}[htb!]
    \centering
    \includegraphics[width=\linewidth]{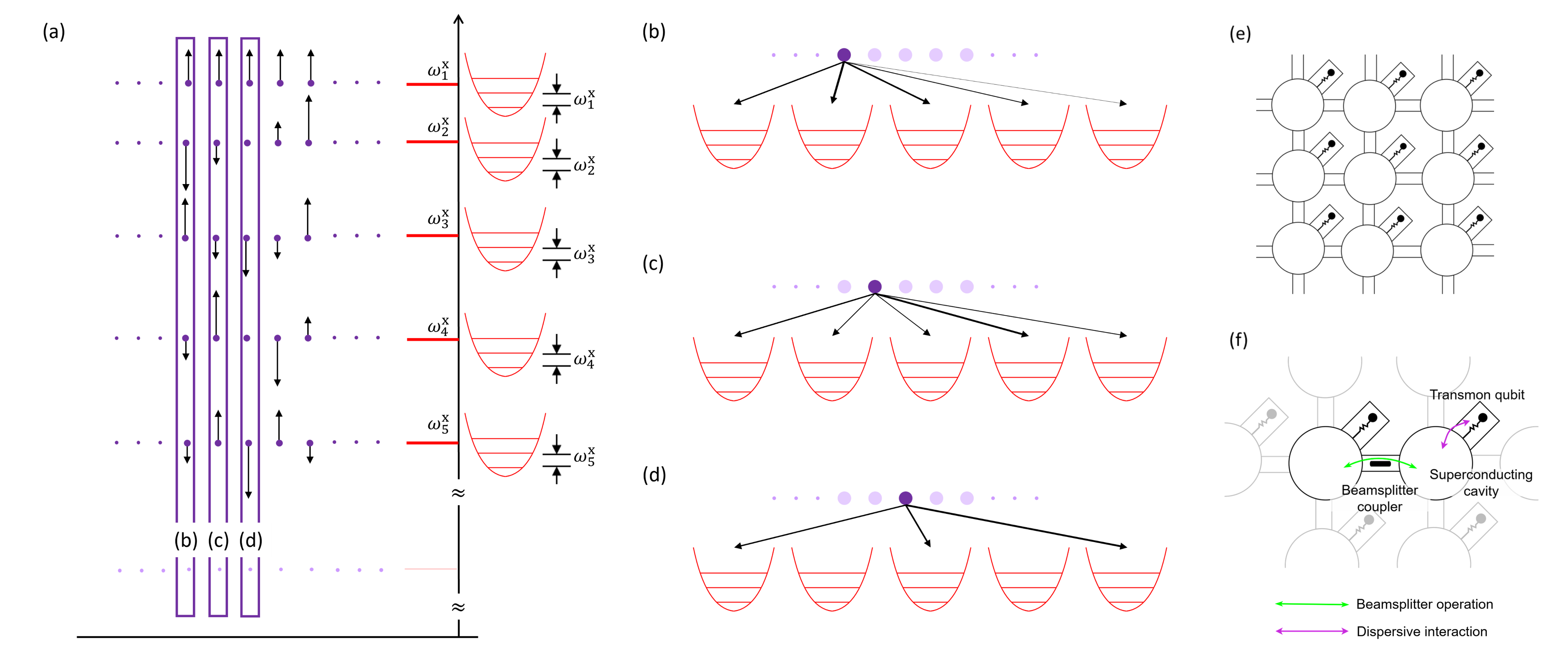}
    \caption{Platform comparison for scalable dual-rail encoding schemes and their dual-rail qubits: (a--d) Trapped-ion system with near all-to-all connectivity. (a) Schematic representation of vibrational mode vectors and corresponding frequencies as harmonic oscillators for the five highest-frequency modes in a linear ion chain. (b-d) Illustration of the coupling strengths between ions and vibrational modes, represented by arrows indicating the relative magnitude of the coupling for each vibrational mode. (e--f) Superconducting platform with nearest-neighbor connectivity. (e) Schematic layout of a 2D superconducting cavity-based system, showing the arrangement of qubits and their connections. (f) Detailed illustration of superconducting cavities coupled to transmon qubits, emphasizing beam splitter couplings and dispersive interactions.}
    \label{fig:comparison}
\end{figure}

When the ions are tightly confined in the trap, all internal qubits can simultaneously interact with vibrational modes, forming all-to-all connectivity (Fig.~\ref{fig:comparison} (a--d)). This connectivity is another advantage of our scheme. As a result, all quantum gate operations on dual-rail qubits require one ancillary internal qubit initialized in the \(\ket{\downarrow}\) state. However, in practical implementations, we must consider the coupling strength between internal qubits and modes (Fig.~\ref{fig:comparison} (b--d)) to minimize errors arising from the off-resonant coupling. To reduce such infidelity, selecting an ion that strongly couples to the desired modes is essential. Consequently, multiple internal qubits may be necessary during dual-rail qubit gate operations to ensure reliable performance. In addition, multiple internal qubits may also be required for parallel gate implementations.

The dominant source of error in modes is heating, which arises from electric field noise and temperature. Heating errors alter the phonon population in the modes, which causes the quantum state of the dual-rail qubit to leave the logical subspace. To address this issue, we propose a circuit for QND measurement of the total phonon number parity in the modes of a dual-rail qubit (see Appendix~\ref{Appendix: Bosonic Qubit Error Detection} for more details). This circuit enables phonon number monitoring without disturbing the quantum state of the dual-rail qubit, provided that no such error occurs. However, performing intermediate measurements in a trapped-ion system remains an open question, as existing measurement methods for internal qubits disrupt the quantum state of the modes. Although we can employ dark state detection to monitor the total phonon number parity in a dual-rail qubit, the quantum information process must be restarted if a parity error is detected.

Another quantum platform proposed for dual-rail encoding is the superconducting platform~\cite{teoh2023dual}. Superconducting platform comprises microwave cavities, beamsplitter couplers, and transmon qubits (Fig.~\ref{fig:comparison} (e--f)). The dual-rail qubit is encoded using two cavity modes and a single photon. The superconducting platform offers inherent scalability, as its components can be engineered through circuit design.

Beamsplitter couplers connect the cavities and apply beamsplitter operations on the associated cavity modes, thereby implementing a single-qubit gate for dual-rail qubits. They are realized through nonlinear coupling between microwave resonators, driven by RF signals such as SNAILs~\cite{chapman2023high} or differentially driven SQUIDs~\cite{lu2023high}.

Transmon qubits are coupled to each cavity to enable nonlinear operations on the cavity modes through dispersive interactions. This interaction is analogous to a vibrational mode and an internal qubit in a trapped-ion system. For state preparation, a single photon is loaded into a cavity using control pulses~\cite{heeres2017implementing} or cavity-transmon sideband drives \cite{elder2020high,wallraff2007sideband,premaratne2017microwave,rosenblum2018cnot}. Dual-rail qubits are measured using a photon number parity operation, followed by transmon readout~\cite{sun2014tracking}. This operation is realized via dispersive interactions between the transmon and the cavity mode, which map photon number parity of the cavity mode onto the transmon.

Beamsplitter couplers and dispersive interactions enable the implementation of the TNP phase gate on cavity modes, and accordingly, the RZZ gate on dual-rail qubits, as described in Sec.~\ref{subsub: TNP phase gate and two-qubit gate}. Therefore, the dual-rail encoding scheme with a superconducting platform can achieve universal quantum computation. However, since beamsplitter couplers determine the connectivity of the cavity modes, the connectivity of the dual-rail qubit system depends on the circuit design.

The dominant source of error for dual-rail qubits in superconducting cavities is photon loss due to cavity relaxation. This error alters the photon population in the cavity modes, causing the quantum state of the dual-rail qubit to leave the logical subspace. Although the nature of this error differs from that in the trapped-ion system, its effect on the dual-rail qubit is similar. In both cases, quantum information is lost due to fluctuations in the boson population.

As in the trapped-ion system, Ref.~\cite{teoh2023dual} proposes a QND measurement scheme to detect such errors by checking the total photon number parity in the cavity modes of the dual-rail qubit. However, unlike the trapped-ion system, the superconducting platform enables transmon readout without disturbing the cavity state, thereby allowing QND measurement. Once the photon number parity is obtained, quantum information processing can proceed, with dual-rail qubits in which such an error has occurred being discarded.

\begin{table}[htb!]
    \centering
    \includegraphics[width=0.8\linewidth]{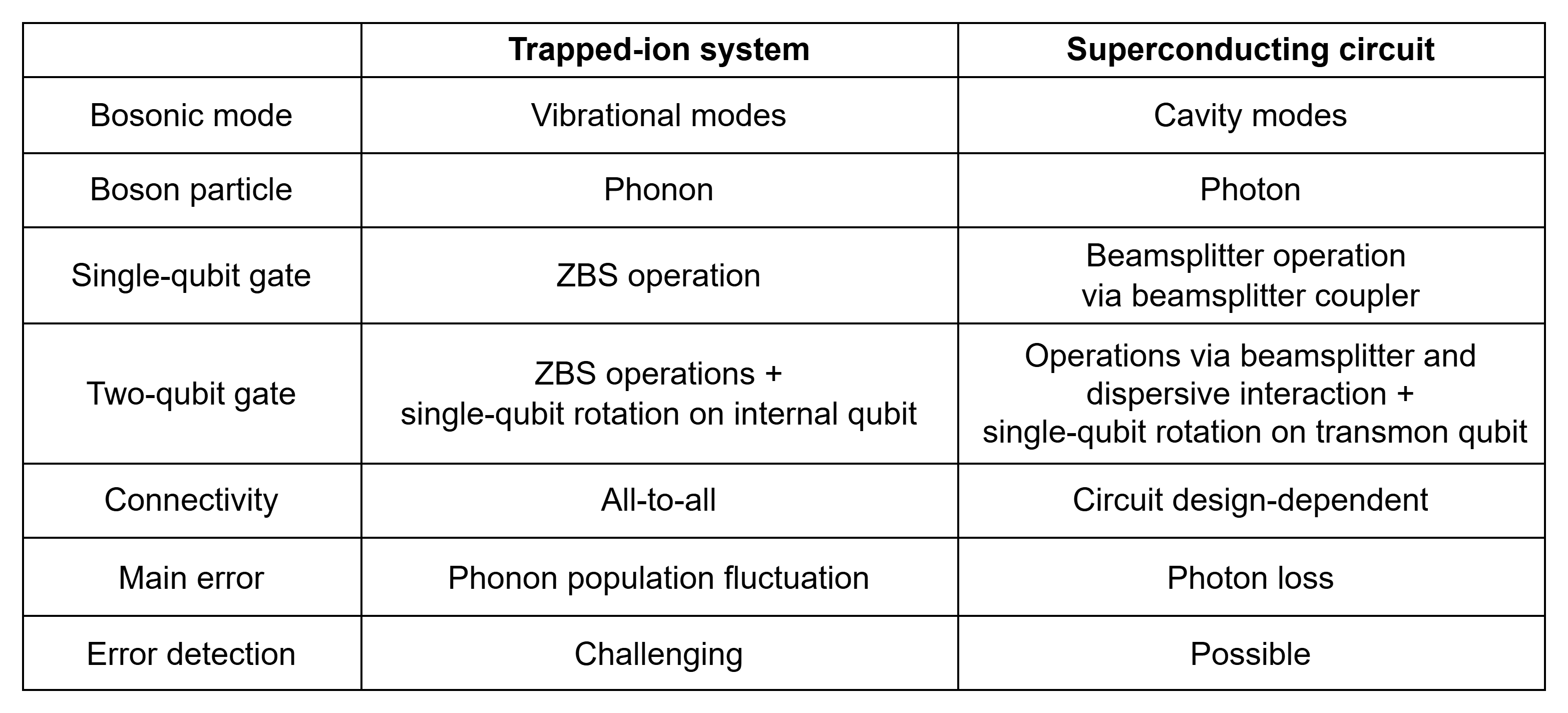}
    \caption{Comparison of Dual-Rail Encoding Schemes in Trapped-Ion Systems and Superconducting Circuits
    }
    \label{tab:my_label}
\end{table}

\section{Logical internal qubit--dual-rail qubit hybrid system}
\label{Sec: hybrid system}

Owing to the all-to-all connectivity, quantum gates acting on dual-rail qubits only require one ancillary internal qubit in the trapped-ion system. Therefore, the other internal qubits can be utilized as additional logical qubits, called logical internal qubits. Hence, in this section, we introduce the logical internal qubit—dual-rail qubit hybrid system, which is constructed by incorporating these logical internal qubits into the dual-rail encoding scheme with the trapped-ion phononic network.

A single-qubit rotation on internal qubits can implement the single-qubit gate acting on logical internal qubits. In addition, existing methods such as the Mølmer–Sørensen gate~\cite{sorensen1999quantum} and the Cirac-Zoller gate~\cite{cirac1995quantum}, enable two-qubit gates between logical internal qubits. Both methods rely on interactions between modes and internal qubits, and the choice of method determines the number of ancillary modes required: the  Mølmer–Sørensen gate requires no ancillary mode, while the Cirac-Zoller gate requires one. Consequently, in this hybrid system, at most two ancillary resources are needed: one internal qubit and, if necessary, one ancillary mode.

Thus, the remaining quantum gate to be implemented in the hybrid system is a two-qubit gate between a logical internal qubit and a dual-rail qubit. In Sec.~\ref{Subsec: internal dual-rail two qubit}, we introduce two methods for implementing such gates: the CBS gate and the \(XX\)-rotation (RXX) gate. Notably, their realization does not require additional ancillary resources from the hybrid system. Furthermore, the controlled-SWAP (CSWAP) gate can be implemented using CBS gates, and its construction is presented in Sec.~\ref{subsec: cswap}.

This hybrid system still preserves the all-to-all connectivity, thus all logical qubits in the hybrid system can be entangled directly. Moreover, the hybrid system can increase the number of logical qubits beyond a traditional quantum computer of the trapped-ion system without adding more ions: for \(N\) ions, we can obtain at most \(N\) dual-rail qubits. Thus, the hybrid system can accommodate nearly twice as many logical qubits as a conventional trapped-ion quantum computer.

In Sec.~\ref{subsec: Toffoli gate}, we propose a method for the efficient implementation of multi-qubit controlled gates in the hybrid system. It is based on the Cirac-Zoller multi-qubit gate protocol~\cite{cirac1995quantum}, which implements a multi-qubit CNOT gate for the logical internal qubits by introducing a third energy level as an auxiliary state. This protocol has been experimentally demonstrated in recent work~\cite{fang2023realization}.

In our method, the auxiliary state of a logical qubit is encoded using an auxiliary mode, thus extending the Cirac-Zoller scheme from the logical internal qubit system to the hybrid system. Furthermore, we generalize our method to implement arbitrary multi-qubit-controlled gates.

Finally, in Sec.~\ref{subsec: application}, we discuss the potential applications of the hybrid system. Although current techniques for measuring internal qubits destroy the quantum information stored in the vibrational modes, making simultaneous measurement of all logical qubits in the hybrid system impossible, specific applications can still exploit the hybrid system's advantages. In particular, the set of dual-rail qubits can serve as an auxiliary register that is not subject to measurement. In addition, the all-to-all connectivity of the hybrid system enables the flexible rearrangement of qubit locations, allowing measurement qubits to be designated as internal qubits without modifying the abstract quantum circuit.

\subsection{Two-qubit gate between the logical internal qubit and the dual-rail qubit with ZBS gate}
\label{Subsec: internal dual-rail two qubit} 

As presented in Sec.~\ref{subsub: ZBS and single-qubit gate}, the ZBS gate applies different beamsplitter operations to the modes depending on the internal qubit state, inherently enabling a two-qubit gate between a logical internal qubit and a dual-rail qubit. This section presents two methods for implementing such two-qubit gates using ZBS gates: the CBS and RXX gates.

The CBS gate is a hybrid gate that applies a beamsplitter operation to the modes conditioned on the internal qubit being in a specific state~\cite{gan2020hybrid}. For an internal qubit \(q\) and two modes \((b_1,b_2)\), the CBS gate is expressed as:
\begin{align}
\label{eqn: CBS}
\hat{U}_{\text{CBS}}(\theta,\phi) = \exp{(i\theta \Op{\uparrow}{\uparrow} \otimes (\Cre_{b_{1}}\Ann_{b_{2}} e^{i\phi} + \Ann_{b_{1}}\Cre_{b_{2}} e^{-i\phi}))},
\end{align}
such that
\begin{gather}
    \hat{U}_{\text{CBS}}(\theta,\phi)\Ket{\downarrow}\Ket{\psi} = \Ket{\downarrow} \Ket{\psi}, \\
    \hat{U}_{\text{CBS}}(\theta,\phi)\Ket{\uparrow}\Ket{\psi} = \Ket{\uparrow}\hat{B}(\theta,\phi)\ket{\psi}.
\end{gather}
Here, \(\Ket{\psi}\) represents the quantum state of modes \((b_1,b_2)\), and \(\hat{B}(\theta,\phi)\) is the beamsplitter operator (Eq.~\ref{Eqn: Beamsplitter}). The CBS gate can also be rewritten as:
\begin{align}
    \label{eqn: CBS_split}
    \hat{U}_{\mathrm{CBS}}(\theta,\phi) = \ket{\downarrow}\bra{\downarrow}\otimes \hat{I} + \ket{\uparrow}\bra{\uparrow}\otimes\hat{B}(\theta,\phi).
\end{align}
As described in Introduction, although the CBS gate has been experimentally demonstrated~\cite{gan2020hybrid}, the experiment was limited to a single-ion trap system, leaving its scalability unproven. Instead, we propose an implementation of the CBS gate using ZBS gates.

\begin{figure}[ht]
    \centering
    \includegraphics[width=0.8\linewidth]{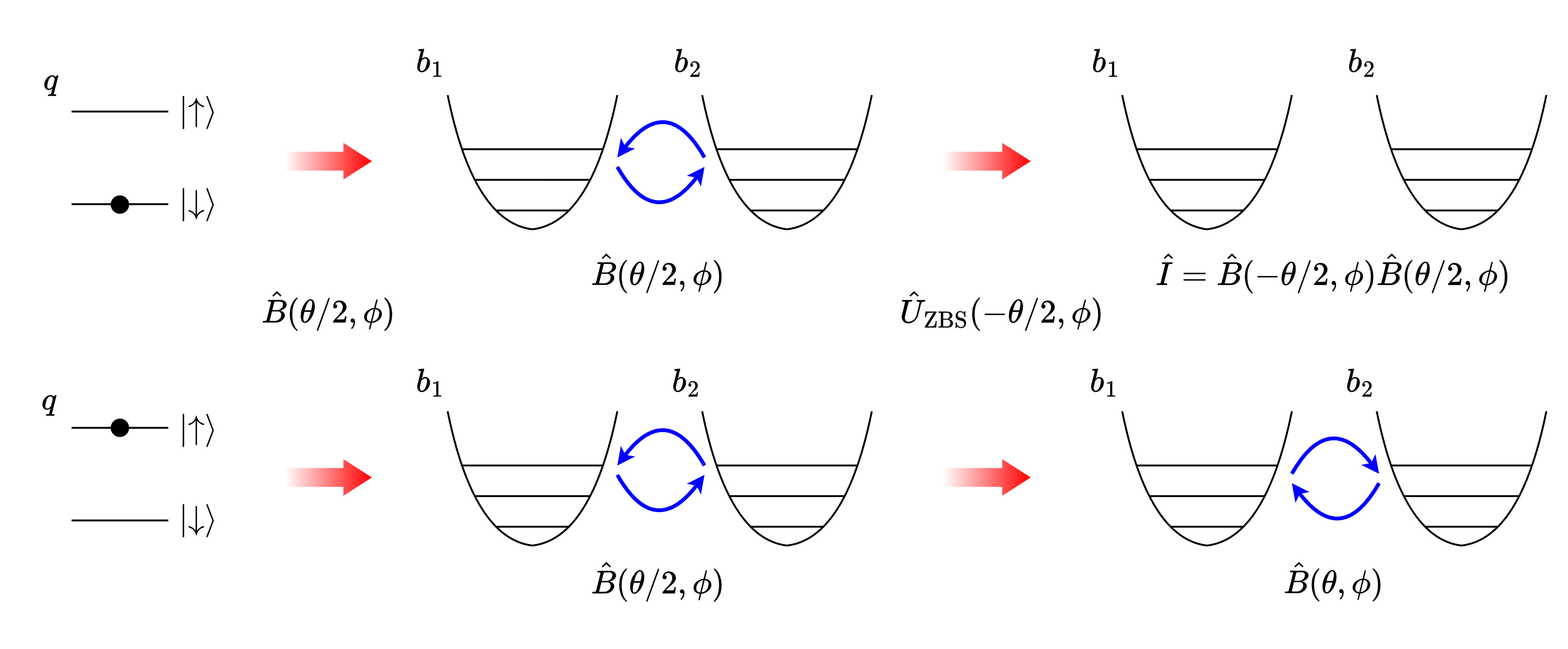}
    \caption{Graphical representation of the CBS gate derived from the ZBS gate, based on the identities \(\hat{I} = \hat{B}(\theta/2,\phi)\hat{B}(-\theta/2,\phi)\) and \(\hat{B}(\theta,\phi) = \hat{B}(\theta/2,\phi)^2\).}
    \label{fig: cbs_figure}
\end{figure}

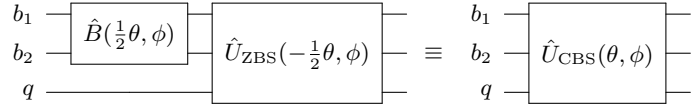
\begin{figure}[ht]
\centering
\[ \Qcircuit @C=1em @R=.7em{        
            \lstick{b_1} &\multigate{1}{\hat{B}(\frac{1}{2}\theta,\phi)} & \multigate{2}{\hat{U}_{\text{ZBS}}(-\frac{1}{2}\theta,\phi)} & \qw&& && \lstick{b_1} & \multigate{2}{\hat{U}_{\text{CBS}}(\theta,\phi)} & \qw \\
            \lstick{b_2}                                                &\ghost{\hat{B}(\frac{1}{2}\theta,\phi)} & \ghost{\hat{U}_{\text{ZBS}}(-\frac{1}{2}\theta,\phi)} & \qw & &\lstick{\equiv} && \lstick{b_2}& \ghost{\hat{U}_{\text{CBS}}(\theta,\phi)} & \qw\\
            \lstick{q}       & \qw & \ghost{\hat{U}_{\text{ZBS}}(-\frac{1}{2}\theta,\phi)} & \qw &&  && \lstick{q} & \ghost{\hat{U}_{\text{CBS}}(\theta,\phi)} & \qw\\ 
            }
        \]
\caption{CBS gate implementation using a ZBS gate and a beamsplitter operator, based on the relation between \(\hat{\sigma}_z\) and \(\Op{1}{1}\).}
\label{Fig: CBS with BS and ZBS}
\end{figure}

Consider the internal qubit parts in the exponents of the ZBS~(Eq.~\eqref{Eqn: ZBS}) and CBS (Eq.~\eqref{eqn: CBS}) gates. The CBS gate contains a projector onto the \(\Ket{\uparrow}\) state, whereas the ZBS gate contains \(\hat{\sigma}_\mathrm{z}\). The relationship between these operators is expressed as \(\hat{\sigma}_\mathrm{z} = \Op{\uparrow}{\uparrow}-\Op{\downarrow}{\downarrow} = \hat{I} - 2\Op{\downarrow}{\downarrow}\). Therefore, the CBS gate can be decomposed into a ZBS gate and a beamsplitter operator~(Figs.~\ref{fig: cbs_figure} and \ref{Fig: CBS with BS and ZBS}):
\begin{gather}\label{Eqn: ZBS to CBS}
        \hat{U}_{\text{CBS}}(\theta,\phi) = \hat{U}_{\text{ZBS}}(-\theta/2,\phi)\hat{B}(\theta/2,\phi).
\end{gather}
The beamsplitter operation in Eq.~\ref{Eqn: ZBS to CBS} is realized using the ZBS gate with an ancillary internal qubit prepared in the \(\ket{\downarrow}\) state, as described in Sec.~\ref{subsub: ZBS and single-qubit gate}.

\begin{figure}[htb!]
    \centering
    \includegraphics[width=0.5\linewidth]{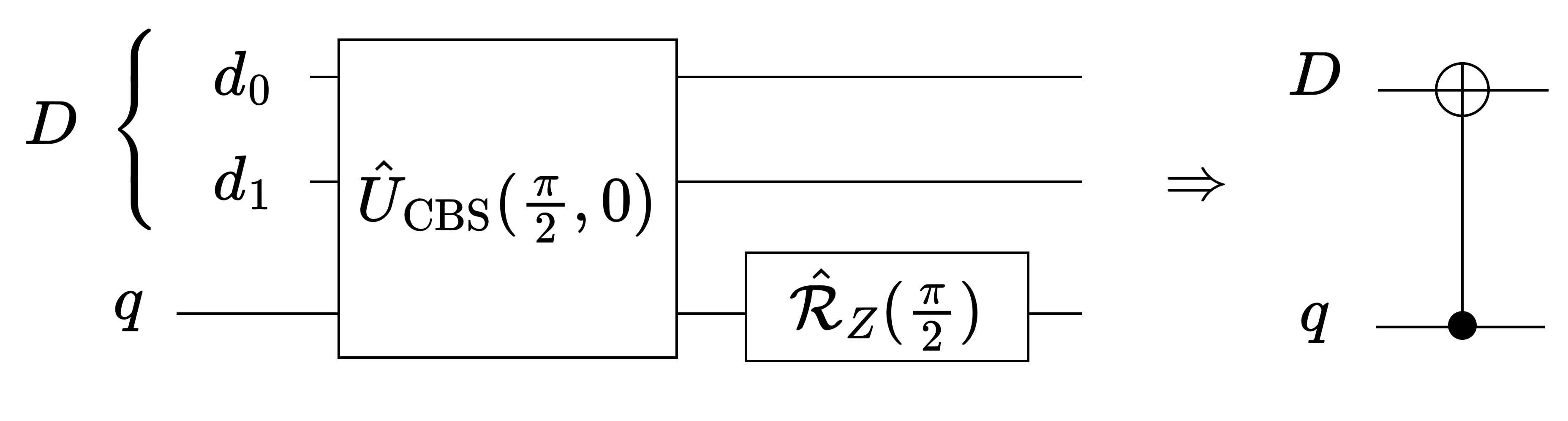}
    \caption{CNOT gate implementation between the logical internal qubit \(q\) and dual-rail qubit \(D\).}
    \label{Fig: CNOT}
\end{figure}

One example of a two-qubit gate between a logical internal qubit and a dual-rail qubit implemented via the CBS gate is the CNOT gate. Specifically, the CBS gate \(\hat{U}_{\mathrm{CBS}}(\pi/2,0)\) from Eq.~\ref{eqn: CBS_split} can be mapped to the logical internal qubit and the dual-rail qubit as follows:
\begin{align}
    \hat{U}_{\mathrm{CBS}}(\pi/2,0) = \ket{\downarrow}\bra{\downarrow}\otimes \hat{I} + \ket{\uparrow}\bra{\uparrow}\otimes \hat{B}(\pi/2,0) \Rightarrow \ket{0}\bra{0}_q\otimes \hat{I}_D + \ket{1}\bra{1}_q\otimes i\hat{X}_D.
\end{align}
The additional phase of \(i\) can be eliminated by applying \(\hat{R}_Z(-\pi/2)\) to the logical internal qubit, which is implemented by the \(\sigma_{\mathrm{z}}\)-axis rotation on internal qubit \(\hat{\mathcal{R}}_Z(\pi/2)\):
\begin{align}
    \hat{\mathcal{R}}_Z(\pi/2)\hat{U}_{\text{CBS}}(\pi/2,0) \Rightarrow  e^{i\pi/4}\left(\Op{0}{0}_{q}\otimes \hat{I}_{D} + \Op{1}{1}_{q} \otimes \hat{X}_{D}\right),
    \label{Eqn: CNOT}
\end{align}
with a global phase of \(e^{i\pi/4}\) (Fig.~\ref{Fig: CNOT}). Here, the logical internal qubit serves as the control qubit, and the dual-rail qubit serves as the target qubit. The roles of the control and target qubits can be interchanged by applying Hadamard gates or \(Y\)-rotation gates to both qubits.

Another two-qubit gate is the RXX gate. For the logical internal qubit \(q\) and the dual-rail qubit \(D\), the RXX gate \(\hat{R}_{XX}(\theta)\) is implemented as follows (Fig.~\ref{Fig: R_xx}, see Appendix~\ref{Appendix: rotation XX} for an explicit derivation):
\begin{align}
    \label{eqn: RXX_}
     \hat{U}_{\mathrm{ZBS}}(\pi/4,0)\hat{\mathcal{R}}_Y(-\theta)\hat{U}_{\mathrm{ZBS}}(-\pi/4,0) \Rightarrow \hat{R}_{XX}(\theta).
\end{align}
Unlike the CBS gate, the RXX gate implementation shown in Eq.~\ref{eqn: RXX_} does not require an ancillary internal qubit.
\begin{figure}[htb!]
    \centering
    \includegraphics[width=0.65\linewidth]{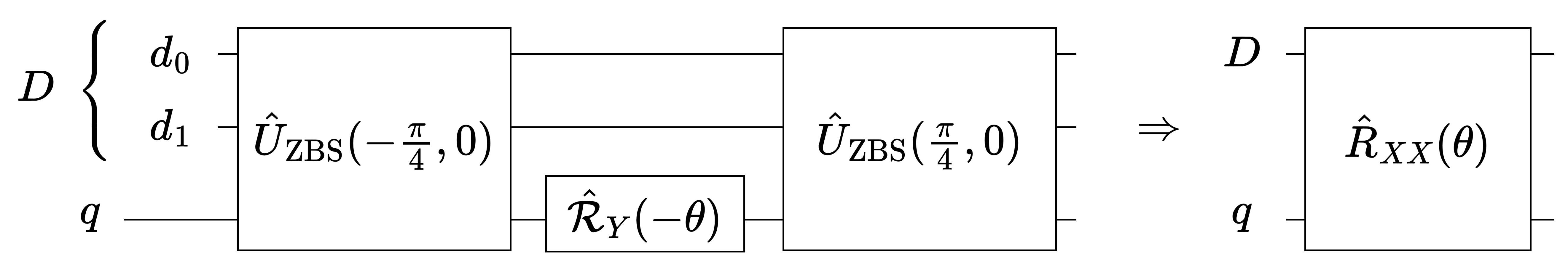}
    \hfill
    \caption{RXX gate implementation between the logical internal qubit \(q\) and dual-rail qubit \(D\).}
    \label{Fig: R_xx}
\end{figure}

\subsection{CSWAP gate using CBS gates}
\label{subsec: cswap}

The CBS gate applies a beamsplitter operation on two modes conditioned on the basis state of the internal qubit. Therefore, a CSWAP gate, in which the logical internal qubit is the control qubit and the dual-rail qubits are the target qubits, can be implemented using two CBS gates and a single-qubit gate on the logical internal qubit.

\begin{figure}[htb!]
    \centering
    \includegraphics[width=0.8\linewidth]{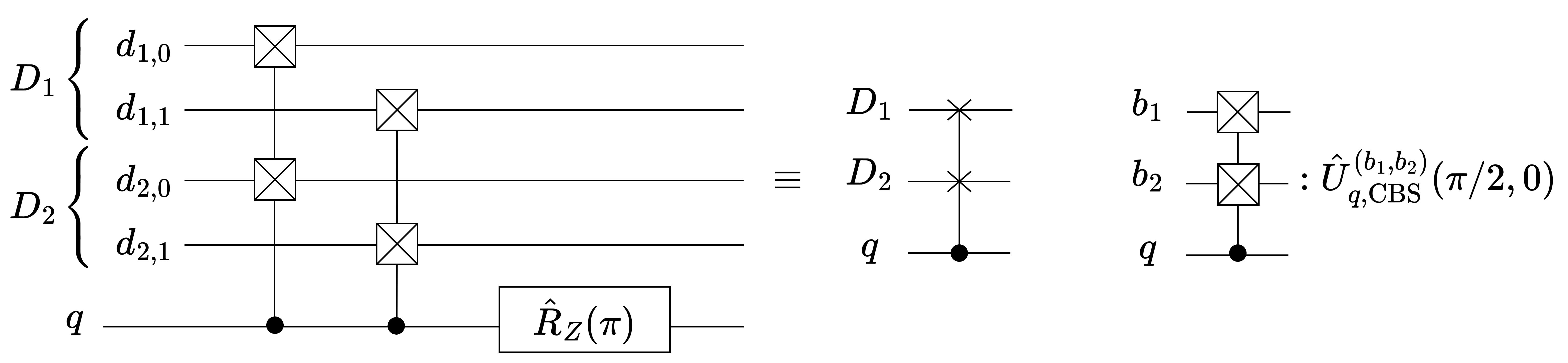}
    \caption{Implementation of CSWAP gate acting on two dual-rail qubits \(D_1\) and \(D_2\), with the logical internal qubit \(q\) as the control. Two CBS gates conditionally swap the quantum states of each dual-rail qubit with a relative phase factor of \(-1\), and it is corrected by applying \(\hat{\mathcal{R}}_Z(-\pi)\) on the internal qubit.}
    \label{fig: cswap gate}
\end{figure}

Consider two dual-rail qubits \(D_1\) and \(D_2\) and an (logical) internal qubit \(q\). The combination of two CBS gates, \(\hat{U}_{\rm{CBS}}^{(d_{1,0},d_{2,0})}(\pi/2,0)\hat{U}_{\rm{CBS}}^{(d_{1,1},d_{2,1})}(\pi/2,0)\), swaps the states of dual-rail qubits with a relative phase of \(-1\), depending on the basis state of the internal qubit \(q\):

\begin{gather}
\hat{U}_{q,\rm{CBS}}^{(d_{1,0},d_{2,0})}(\pi/2,0)\,\hat{U}_{\rm{CBS}}^{(d_{1,1},d_{2,1})}(\pi/2,0)\ket{\downarrow}\ket{\psi_1}_{D_1}\ket{\psi_2}_{D_2} =\ket{\downarrow}\ket{\psi_1}_{D_1}\ket{\psi_2}_{D_2} \Rightarrow \ket{0}_q\ket{\psi_1}_{D_1}\ket{\psi_2}_{D_2},\\
\hat{U}_{q,\rm{CBS}}^{(d_{1,0},d_{2,0})}(\pi/2,0)\,\hat{U}_{\rm{CBS}}^{(d_{1,1},d_{2,1})}(\pi/2,0)\ket{\uparrow}\ket{\psi_1}_{D_1}\ket{\psi_2}_{D_2} =-\ket{\uparrow}\ket{\psi_2}_{D_1}\ket{\psi_1}_{D_2}\Rightarrow -\ket{1}_q\ket{\psi_2}_{D_1}\ket{\psi_1}_{D_2}.
\end{gather}
For clarity, we explicitly denote the bosonic modes acted upon by the CBS gate using superscripts. By subsequently applying a \({\hat{\sigma}_{\mathrm{z}}}\)-axis rotation \(\hat{\mathcal{R}}_Z(-\pi)\) to the internal qubit, the CSWAP gate is realized, up to a global phase of \(e^{i\pi/2}\) (Fig.~\ref{fig: cswap gate}).

In general, the CSWAP gate between two quantum states, each composed of \(N\) dual-rail qubits, can be implemented using \(2N\) CBS gates and at most one \(\hat{\sigma}_{\mathrm{z}}\)-axis rotation on the internal qubit \(\hat{\mathcal{R}}_Z(-\pi)\). Let the first and the second \(N\) dual-rail qubits be denoted as \(D_1,\cdots,D_N\) and \(D_{N+1},\cdots,D_{2N}\), respectively. Then, the product of CBS gates pairs, given by 
\begin{align}
    \prod_{i=1}^{N}(\hat{U}_{\mathrm{CBS}}^{(d_{i,0},d_{N+i,0})}(\pi/2,0)\hat{U}_{\mathrm{CBS}}^{(d_{i,1},d_{N+i,1})}(\pi/2,0),
\end{align}
swaps two quantum states of \(N\) dual-rail qubits with a relative phase of \((-1)^N\), only when the internal qubit is in the state \(\ket{\uparrow}\). Therefore, by subsequently applying \(\hat{R}_{Z}(-\pi)\) to the internal qubit for odd \(N\), the CSWAP gate between two quantum states of \(N\) dual-rail qubits can be realized.

\subsection{Efficient implementation of multi-qubit controlled gates in the hybrid system}

\label{subsec: Toffoli gate}

In this section, we propose a method for implementing multi-qubit controlled gates in the hybrid system. This approach builds on the Cirac--Zoller multi-qubit gate protocol introduced in Ref.~\cite{fang2023realization}, enabling the realization of \(K\)-qubit CNOT (\(K\)-CNOT) gates for logical internal qubits with \(\mathcal{O}(K)\) operations in a trapped-ion system. Cirac-Zoller approach introduces an auxiliary state within the internal qubit and the RSB transition to couple the ground state and the auxiliary state of the internal qubit to the center-of-mass (COM) mode (denoted as \(b_{\mathrm{COM}}\)). In this protocol, the COM mode is initially in the vacuum state \(\ket{\underline{0}}\).

The universality of quantum computation ensures that any quantum gate can be decomposed into single-qubit and CNOT gates~\cite{deutsch1995universality}. However, such decompositions typically require an exponentially large number of operations, which motivates the direct implementation of multi-qubit gates. In trapped-ion systems, several schemes for such multi-qubit operations have been proposed~\cite{cirac1995quantum, wang2001multibit,levine2019parallel,katz2022n,fang2023realization}.

Recently, the Cirac-Zoller method was experimentally demonstrated in Ref.~\cite{fang2023realization}. The demonstration was performed using a system of \(^{171}\)Yb\(^{+}\) ions confined in a microfabricated linear Paul trap~\cite{revelle2020phoenix,wang2020high}. The internal qubit is encoded in two hyperfine clock states of the \(^{2}\)S\(_{1/2}\) manifold: \(\ket{\downarrow} \equiv \ket{F=0,m_F=0}\) and \( \ket{\uparrow} \equiv \ket{F=1,m_F=0}\), which define the logical basis states: \(\ket{0}\Rightarrow \ket{\downarrow}\) and \(\ket{1}\Rightarrow\ket{\uparrow}\). In addition, one of the Zeeman levels in the \(^{2}\)S\(_{1/2}\) ground state \(\ket{F=1,m_F=\pm 1}\), is employed as the auxiliary state, \(\ket{2}\), of the logical internal qubit.

The RSB transition \(\hat{U}_{\mathrm{RSB}}(\theta)\) (Eq.~\ref{eqn: transition}) on the logical internal qubit and the mode can be expressed as:
\begin{align}
    \label{eqn: RSB encode}
    \hat{U}_{\rm{RSB}}(\theta) \Rightarrow \exp\left(i\frac{\theta}{2}\left(\ket{1}\bra{0}\hat{a} + \ket{0}\bra{1}\hat{a}^{\dagger}\right) \right).
\end{align}
The action of \(\hat{U}_{\rm{RSB}}(\pi)\) on the logical internal qubit and the COM mode yields:
\begin{gather}
\label{eqn: RSB_1}
\Ket{0}\Ket{\underline{1}} \xrightarrow{\hat{U}_{\rm{RSB}}(\pi)} -i\Ket{1}\Ket{\underline{0}}  \xrightarrow{\hat{U}_{\rm{RSB}}(\pi)} -\Ket{0}\Ket{\underline{1}}, \\
\Ket{0}\Ket{\underline{0}} \xrightarrow{\hat{U}_{\rm{RSB}}(\pi)} \Ket{0}\Ket{\underline{0}}, \\
\label{eqn: RSB_3}
\Ket{1}\Ket{\underline{1}}\xrightarrow{\hat{U}_{\rm{RSB}}(\pi)} -\Ket{1}\Ket{\underline{1}}.
\end{gather}

\begin{figure}[htb!]
    \centering
    \includegraphics[width=0.7
    \linewidth]{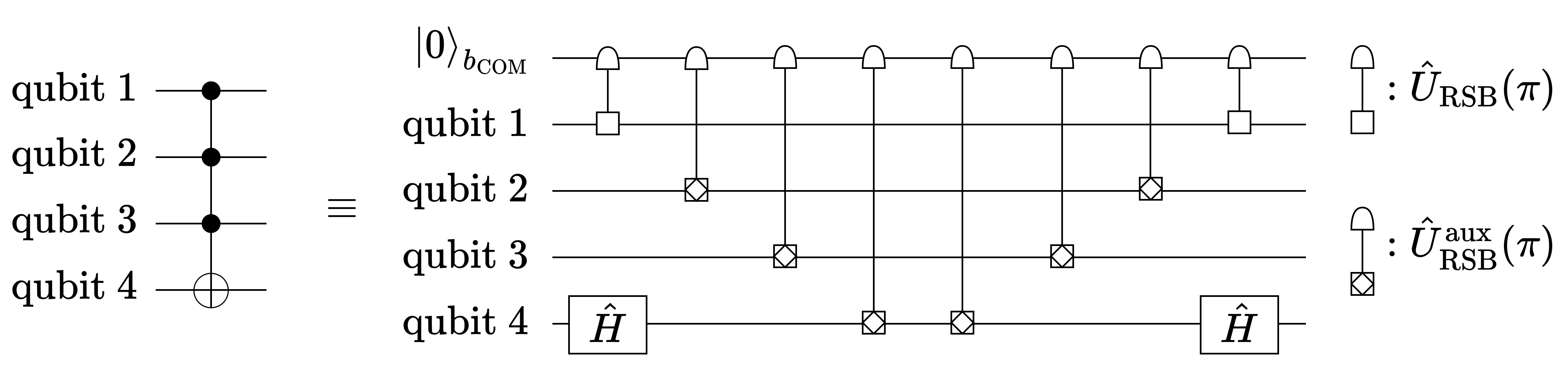}
    \caption{\(K\)-CNOT gate implementation in a trapped-ion system using \(\hat{U}_{\rm{RSB}}(\pi)\) and \(\hat{U}_{\rm{RSB}}^{\rm{aux}}(\pi)\).}
    \label{fig:N-Toffoli}
\end{figure}

RSB transition can also couple \(\ket{0}\) and \(\ket{2}\) of the logical internal qubit to the mode. We denote such RSB transition as \(\hat{U}_{\rm{RSB}}^{\rm{aux}}(\theta)\):
\begin{align}
    \hat{U}_{\rm{RSB}}^{\rm{aux}}(\theta) \Rightarrow \exp\left(i\frac{\theta}{2}\left(\ket{2}\bra{0}\hat{a} + \ket{0}\bra{2}\hat{a}^{\dagger}\right) \right).
\end{align}
The action of \(\hat{U}_{\rm{RSB}}^{\rm{aux}}(\pi)\) on the logical internal qubit and the COM mode is:
\begin{gather}
\label{eqn:RSB_aux_1}
\Ket{0}\Ket{\underline{1}} \xrightarrow{\hat{U}_{\rm{RSB}}^{\rm{aux}}(\pi)} -i\Ket{2}\Ket{\underline{0}}  \xrightarrow{\hat{U}_{\rm{RSB}}^{\rm{aux}}(\pi)} -\Ket{0}\Ket{\underline{1}}, \\
\label{eqn:RSB_aux_2}
\Ket{0}\Ket{\underline{0}} \xrightarrow{\hat{U}_{\rm{RSB}}^{\rm{aux}}(\pi)} \Ket{0}\Ket{\underline{0}},\\
\label{eqn:RSB_aux_3}
\ket{1}\ket{\underline{0}} \xrightarrow{\hat{U}_{\rm{RSB}}^{\rm{aux}}(\pi)} \ket{1}\ket{\underline{0}}, \\
\label{eqn:RSB_aux_4}
\ket{1}\ket{\underline{1}} \xrightarrow{\hat{U}_{\rm{RSB}}^{\rm{aux}}(\pi)} \ket{1}\ket{\underline{1}}.
\end{gather}
The \(K\)-CNOT gate on internal qubits can be implemented using \(\mathcal{O}(K)\) RSB transitions of the logical internal qubits (Fig.~\ref{fig:N-Toffoli}). Note that the COM mode contains at most a single phonon throughout this process.

Our method encodes the auxiliary state \(\ket{2}\) of the logical internal qubit using an auxiliary mode instead of a Zeeman state of the ion. For an internal qubit \(q\), we introduce an auxiliary mode \(b_q\) to encode the auxiliary state. The basis states of the logical internal qubit are encoded as follows:
\begin{align}
    \label{eqn:encode_int_1}
    \ket{\downarrow}_q\ket{\underline{0}}_{b_q} \Rightarrow \ket{0}_q,\\
    \ket{\uparrow}_q\ket{\underline{0}}_{b_q} \Rightarrow \ket{1}_q,\\
    \label{eqn:encode_int_3}
    \ket{\downarrow}_q\ket{\underline{1}}_{b_q} \Rightarrow \ket{2}_q.
\end{align}

The RSB transition that couples \(\ket{0}\) and \(\ket{1}\) of the logical internal qubit to the COM mode operates in the same way as the original method. However, the auxiliary state of the logical internal qubit is now encoded in the auxiliary mode, it is necessary to find a quantum operation that mimics the effect of \(\hat{U}_{\rm{RSB}}^{\rm{aux}}(\pi)\) described in Eqs.~\ref{eqn:RSB_aux_1}-\ref{eqn:RSB_aux_4}:
\begin{gather}
\ket{\downarrow}_q\ket{\underline{0}}_{b_q}\ket{\underline{1}}_{b_{\rm{COM}}} \rightarrow -i\ket{\downarrow}_q\ket{\underline{1}}_{b_q}\ket{\underline{0}}_{b_{\rm{COM}}} \rightarrow -\ket{\downarrow}_q\ket{\underline{0}}_{b_q}\ket{\underline{1}}_{b_{\rm{COM}}}, \\
\ket{\downarrow}_q\ket{\underline{0}}_{b_q}\ket{\underline{0}}_{b_{\rm{COM}}} \rightarrow \ket{\downarrow}_q\ket{\underline{0}}_{b_q}\ket{\underline{0}}_{b_{\rm{COM}}},\\
\ket{\uparrow}_q\ket{\underline{0}}_{b_q}\ket{\underline{0}}_{b_{\rm{COM}}} \rightarrow \ket{\uparrow}_q\ket{\underline{0}}_{b_q}\ket{\underline{0}}_{b_{\rm{COM}}}, \\
\ket{\uparrow}_q\ket{\underline{0}}_{b_q}\ket{\underline{1}}_{b_{\rm{COM}}} \rightarrow \ket{\uparrow}_q\ket{\underline{0}}_{b_q}\ket{\underline{1}}_{b_{\rm{COM}}}.
\end{gather}
It is nothing but the CBS gate, which applies a beamsplitter operation between \(b_q\) and \(b_{\rm{COM}}\), conditioned on the internal qubit \(q\) being in the \(\ket{\downarrow}\) state. The corresponding CBS gate can be realized using Eq.~\eqref{eqn: CBS}, where the parameter in the ZBS gate is adjusted from \( -\theta\) to \(\theta\). Therefore, \(\hat{U}_{\rm{RSB}}^{\rm{aux}}(\pi)\) for Eqs.~\ref{eqn:RSB_aux_1}-\ref{eqn:RSB_aux_4} can be replaced as:
\begin{align}
\label{eqn: RSB_aux}
    \hat{U}_\mathrm{RSB}^{\mathrm{aux}}(\pi) \rightarrow  \hat{U}_{\mathrm{ZBS}}(-\pi/4,0)\hat{B}(-\pi/4,0).
\end{align}

Since the CBS gate is implemented using two ZBS gates, the \(K\)-CNOT gate in our method requires \(K\) ancillary modes (one for the COM mode and \(K-1\) for the auxiliary state of the logical internal qubit), one ancillary internal qubit for applying the beamsplitter operations, and \(\mathcal{O}(K)\) operations in total. Although the CBS gate in Eq.~\ref{eqn: RSB_aux} reproduces the effect of \(\hat{R}_{\rm{RSB}}^{\rm{aux}}(\theta)\) only when the COM mode is in the Fock state \(\ket{\underline{0}}\) or \(\ket{\underline{1}}\), this condition is always satisfied in our implementation since the COM mode contains at most one phonon. Therefore, we denote this operation as \(\hat{U}_{\rm{RSB}}^{\rm{aux}}(\theta)\) on the logical internal qubit to ensure consistency with the notation used in the quantum circuit diagram shown in Fig.~\ref{fig:N-Toffoli}.

We can extend the \(K\)-CNOT gate method to dual-rail qubits, enabling its realization within the hybrid system. The auxiliary state of a dual-rail qubit can also be encoded using an auxiliary mode, analogous to the logical internal qubit. Let \(b_D\) be an auxiliary mode of the dual-rail qubit \(D\). The basis states of the dual-rail qubit are encoded as follows:
\begin{align}
    \ket{\underline{1}}_{d_0}\ket{\underline{0}}_{d_1}\ket{\underline{0}}_{b_D} \Rightarrow \ket{0}_D,\\
    \ket{\underline{0}}_{d_0}\ket{\underline{1}}_{d_1}\ket{\underline{0}}_{b_D} \Rightarrow \ket{1}_D,\\
    \ket{\underline{1}}_{d_0}\ket{\underline{0}}_{d_1}\ket{\underline{1}}_{b_D} \Rightarrow \ket{2}_D.
\end{align}

For the \(K\)-CNOT gate implementation, we must implement two quantum operations acting on the dual-rail qubit: the first operation must perform the state transformations described in Eqs.~\ref{eqn: RSB_1}-\ref{eqn: RSB_3}:
\begin{gather} 
\label{eqn:rsb_dual_1}
\Ket{\underline{0}}_{d_0}\ket{\underline{1}}_{d_1}\ket{\underline{0}}_{b_D}\Ket{\underline{0}}_{b_{\text{COM}}} \xrightarrow{} -i\Ket{\underline{1}}_{d_0}\ket{\underline{0}}_{d_1}\ket{\underline{0}}_{b_D}\Ket{\underline{0}}_{b_{\text{COM}}} \xrightarrow{} -\Ket{\underline{0}}_{d_0}\ket{\underline{1}}_{d_1}\ket{\underline{0}}_{b_D}\Ket{\underline{0}}_{b_{\text{COM}}}, \\
    \Ket{\underline{1}}_{d_0}\ket{\underline{0}}_{d_1}\ket{\underline{0}}_{b_D}\Ket{\underline{0}}_{b_{\text{COM}}} \xrightarrow{} \Ket{\underline{1}}_{d_0}\ket{\underline{0}}_{d_1}\ket{\underline{0}}_{b_D}\Ket{\underline{0}}_{b_{\text{COM}}},\\
\label{eqn:rsb_dual_3}    \Ket{\underline{0}}_{d_0}\ket{\underline{1}}_{d_1}\ket{\underline{0}}_{b_D}\Ket{\underline{1}}_{b_{\text{COM}}} \rightarrow -\Ket{\underline{0}}_{d_0}\ket{\underline{1}}_{d_1}\ket{\underline{0}}_{b_D}\Ket{\underline{1}}_{b_{\text{COM}}},
\end{gather}
and the second operation must reproduce the state transformations described in Eqs.~\ref{eqn:RSB_aux_1}-\ref{eqn:RSB_aux_4}:
\begin{gather}
\label{eqn:rsb_dual_aux_1}    \Ket{\underline{1}}_{d_0}\ket{\underline{0}}_{d_1}\ket{\underline{0}}_{b_D}\Ket{\underline{1}}_{b_{\text{COM}}}     \rightarrow -i\Ket{\underline{1}}_{d_0}\ket{\underline{0}}_{d_1}\ket{\underline{1}}_{b_D}\Ket{\underline{0}}_{b_{\text{COM}}} \rightarrow -\Ket{\underline{1}}_{d_0}\ket{\underline{0}}_{d_1}\ket{\underline{0}}_{b_D}\Ket{\underline{1}}_{b_{\text{COM}}},\\
    \Ket{\underline{1}}_{d_0}\ket{\underline{0}}_{d_1}\ket{\underline{0}}_{b_D}\Ket{\underline{0}}_{b_{\text{COM}}} \rightarrow \Ket{\underline{1}}_{d_0}\ket{\underline{0}}_{d_1}\ket{\underline{0}}_{b_D}\Ket{\underline{0}}_{b_{\text{COM}}},\\
    \Ket{\underline{0}}_{d_0}\ket{\underline{1}}_{d_1}\ket{\underline{0}}_{b_D}\Ket{\underline{0}}_{b_{\text{COM}}} \rightarrow \Ket{\underline{0}}_{d_0}\ket{\underline{1}}_{d_1}\ket{\underline{0}}_{b_D}\Ket{\underline{0}}_{b_{\text{COM}}},\\
\label{eqn:rsb_dual_aux_4}    \Ket{\underline{0}}_{d_0}\ket{\underline{1}}_{d_1}\ket{\underline{0}}_{b_D}\Ket{\underline{1}}_{b_{\text{COM}}} \rightarrow \Ket{\underline{0}}_{d_0}\ket{\underline{1}}_{d_1}\ket{\underline{0}}_{b_D}\Ket{\underline{1}}_{b_{\text{COM}}}.
\end{gather}

The central idea is to introduce an ancillary internal qubit \(q\), initialized in the state \(\ket{\downarrow}\), and to exchange the quantum state of two modes \(d_0\) and \(d_1\) with that of the internal qubit. It is based on the CNOT gate between the dual-rail qubit and the logical internal qubit (Eq.~\ref{Eqn: CNOT}). Since the logical qubit \(D\) is a composite system consisting of the conventional dual-rail qubit \((d_0,d_1)\) and the auxiliary mode \(b_D\), The state exchange enables the internal qubit \(q\) and the auxiliary mode \(b_D\) to represent the logical qubit including its auxiliary basis state. Consequently, the desired operations can be implemented using the same operations used for the logical internal qubit.

\begin{figure}
    \centering
    \includegraphics[width=1.0\linewidth]{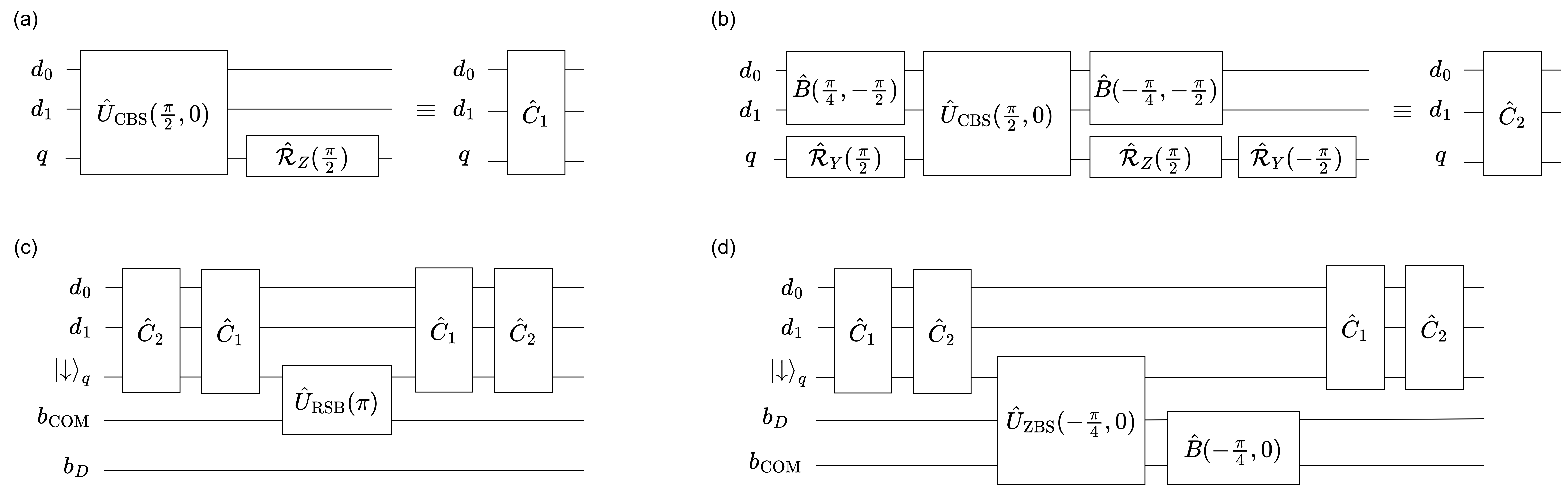}
    \caption{Quantum circuits for coupling a dual-rail qubit to the COM mode for implementing a \(K\)-CNOT gate in the hybrid system. An additional internal qubit \(q\) is introduced to exchange the quantum state of the modes \(d_0\) and \(d_1\) with that of the internal qubit, enabling the application of operations that couple a logical internal qubit to the COM mode.(a--b) Quantum circuits implementing CNOT gates between the conventional dual-rail qubit composed of modes\((d_0,d_1)\) and the logical qubit \(q\). (a) CNOT gate in which the logical internal qubit is the control and the dual-rail qubit is the target, denoted by \(\hat{C_1}\). (b) CNOT gate with reversed roles, where the dual-rail qubit is the control and the logical internal qubit is the target, denoted by \(\hat{C}_2\). (c--d) Quantum circuits are used to couple the dual-rail qubit to the COM mode. These are realized by first exchanging the quantum state of the modes \(d_0\) and \(d_1\) with that of the intneral qubit initialized in \(\ket{\downarrow}\), applying the coupling operations between the logical internal qubit with the COM mode, and then re-exchanging the state of \(d_0\), \(d_1\), and \(q\). (c) Coupling of \(\ket{0}\) and \(\ket{1}\) of the dual-rail qubit to the COM mode (d) Coupling of \(\ket{0}\) and \(\ket{2}\) of the dual-rail qubit to the COM mode.}
    \label{fig:exchange}
\end{figure}

Fig.~\ref{fig:exchange} (a) and (b) illustrate quantum circuits that implement CNOT gates between a logical internal qubit \(q\) and a dual-rail qubit \((d_0,d_1)\). Fig.~\ref{fig:exchange} (a), denoted by \(\hat{C}_1\), is identical to the circuit shown in Fig.~\ref{Fig: CNOT}, and it implements a CNOT gate in which the logical internal qubit acts as the control and the dual-rail qubit as the target. In contrast, Fig.~\ref{fig:exchange} (b), denoted by \(\hat{C}_2\), implements a CNOT gate wth reversed roles: the dual-rail qubit acts as the control and the logical internal qubit as the target.
The composite operations \(\hat{C}_1\hat{C}_2\) and \(\hat{C}_2\hat{C}_1\) effectively exchange the quantum state between the logical internal qubit \(q\) and the dual-rail qubit \((d_0,d_1)\), conditioned on the logical internal qubit or dual-rail qubit being in the \(\ket{0}\) state, respectively. These transformations can be described in terms of the physical modes \(d_0,d_1\) and the internal qubit \(q\) as follows:
\begin{align}
    \hat{C}_1\hat{C}_2\left(\alpha\ket{\underline{1}}\ket{\underline{0}} + \beta\ket{\underline{0}}\ket{\underline{1}} \right)\ket{\downarrow} = \ket{\underline{1}}\ket{\underline{0}}\left(\alpha\ket{\downarrow}+\beta\ket{\uparrow} \right), \\
    \hat{C_2}\hat{C_1}\ket{\underline{1}}\ket{\underline{0}}\left(\alpha\ket{\downarrow}+\beta\ket{\uparrow} \right) = \left(\alpha\ket{\underline{1}}\ket{\underline{0}} + \beta\ket{\underline{0}}\ket{\underline{1}} \right)\ket{\downarrow},
\end{align}
with a global phase of \(e^{i\pi/2}\) that can be ignored.

Therefore, when the logical qubit \(D\) is given and the internal qubit \(q\) is initialized in the state \(\ket{\downarrow}\), \(\hat{C_1}\hat{C_2}\) exchanges the quantum state of the modes \(d_0\) and \(d_1\) with that of the internal qubit \(q\). As a result, the internal qubit \(q\) and the auxiliary mode \(b_D\) can jointly encode the logical internal qubit, including the auxiliary state in Eqs.~\ref{eqn:encode_int_1}-\ref{eqn:encode_int_3}, while the two modes \(d_0\) and \(d_1\) transition to the state \(\ket{\underline{1}}\ket{\underline{0}}\). 

By applying the \(\hat{U}_{\rm{RSB}}(\pi)\) followed by \(\hat{C_1}\hat{C_2}\) (Fig.~\ref{fig:exchange} (c)), the internal qubit \(q\) is returned to the state \(\ket{\downarrow}\) and the resulting transformation corresponds to Eqs.~\ref{eqn:rsb_dual_1}-\ref{eqn:rsb_dual_3}. If \(\hat{U}_{\rm{RSB}}(\pi)\) is replaced with the CBS gate described in Eq.~\ref{eqn: RSB_aux} (Fig.~\ref{fig:exchange} (d)), the transformation instead corresponds to Eqs.~\ref{eqn:rsb_dual_aux_1}-\ref{eqn:rsb_dual_aux_4}. Both require only a constant number of operations, \(\mathcal{O}(1)\). To maintain consistency with the notation used in the quantum circuit diagram shown in Fig.~\ref{fig:N-Toffoli}, we denote the first and the second operations \(\hat{U}_{\rm{RSB}}(\pi)\) and \(\hat{U}_{\rm{RSB}}^{\rm{aux}}(\pi)\), respectively.

Consequently, our method enables the implementation of the \(K\)-CNOT gate in the hybrid system using \(\mathcal{O}(K)\) operations, \(K\) ancillary modes (one for the COM mode and the remaining \(K-1\) for encoding the auxiliary state of the logical qubit), and at most two additional internal qubits (one for performing beamsplitter operation and the other for coupling between the dual-rail qubit and the COM mode). However, since the inclusion of ancillary resources reduces the number of available qubits in the hybrid system, choosing an appropriate value of \(K\) is crucial.

\begin{figure}[htb!]
    \centering
    \includegraphics[width=0.70\linewidth]{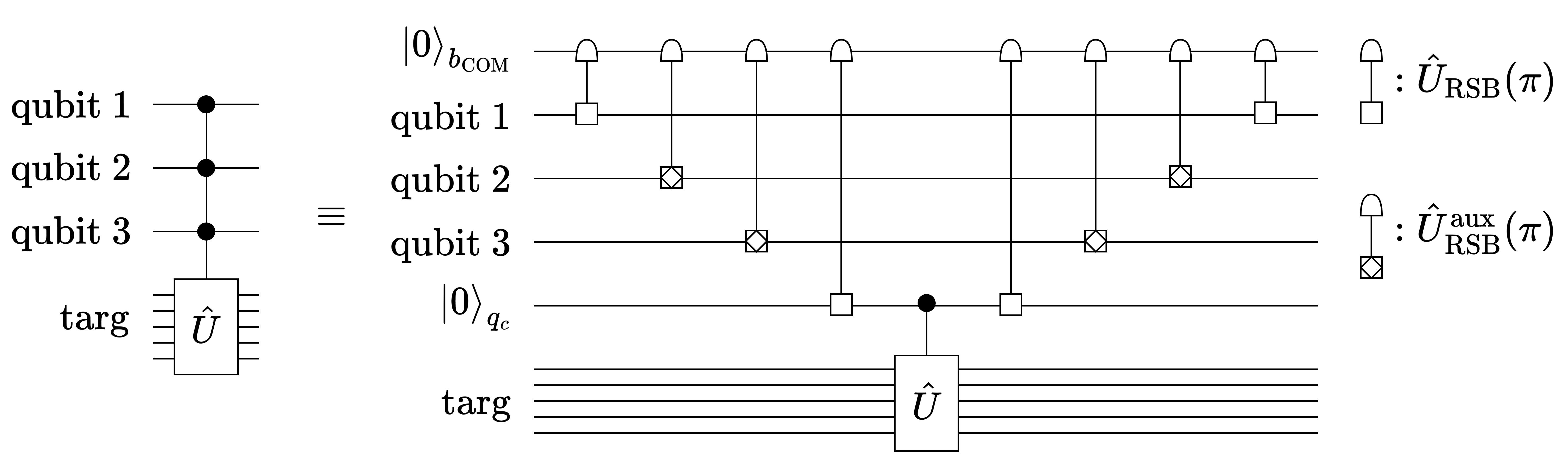}
    \caption{Quantum circuit for implementing a multi-qubit controlled gate using RSB transitions. The presence of a phonon in the COM mode \(b_{\mathrm{COM}}\) after the initial sequence of \(K\) RSB transitions indicates that all control qubits are in the \(\ket{1}\) state. This information is transferred to an auxiliary internal qubit \(q_c\) via an RSB transition. The application of a single-qubit controlled gate with \(q_c\) as the control is thus equivalent to performing the multi-qubit controlled gate. Finally, a second sequence of \(K+1\) RSB transitions restores the states of control qubits and the COM mode, completing the implementation.}
    \label{fig: general multi}
\end{figure}

The \(K\)-CNOT gate can be generalized to an arbitrary multi-qubit controlled gate. In the construction of the \(K\)-CNOT gate, the flipping of the target state depends on the presence of a phonon in the COM mode. In Fig.~\ref{fig:N-Toffoli}, after applying a total of \(K\) unitaries, consisting of \(\hat{U}_{\rm{RSB}}(\pi)\) and \(\hat{U}_{\rm{RSB}}^{\rm{aux}}(\pi)\), the presence of a phonon in the COM mode indicates that all control qubits are in the \(\ket{1}\) state. This condition can be transferred to an additional logical internal qubit \(q_c\), initialized in the \(\ket{0}_{q_c}\) state, by performing an RSB transition \(\hat{U}_{\rm{RSB}}(\pi)\). As a result, \(q_c\) becomes \(-i\ket{1}\) if all control qubits are \(\ket{1}\), and remains \(\ket{0}\) otherwise.

Therefore, an multi-qubit controlled gate can be implemented using a total of \(2K+2\) unitaries of the form \(\hat{U}_{\rm{RSB}}(\pi)\) or \(\hat{U}_{\rm{RSB}}^{\rm{aux}}(\pi)\), together with a single-qubit controlled gate in which the logical interal qubit \(q_c\) act as the control. Specifically, the first \(K+1\) unitaries store the control condition in \(q_c\), the single-qubit controlled gate is applied with \(q_c\) as the control, and the final \(K+1\) unitaries restore the states of control qubits, and return \(q_c\) to \(\ket{0}\) (Fig.~\ref{fig: general multi}). 

\begin{figure}[htb!]
    \centering
    \includegraphics[width=0.70\linewidth]{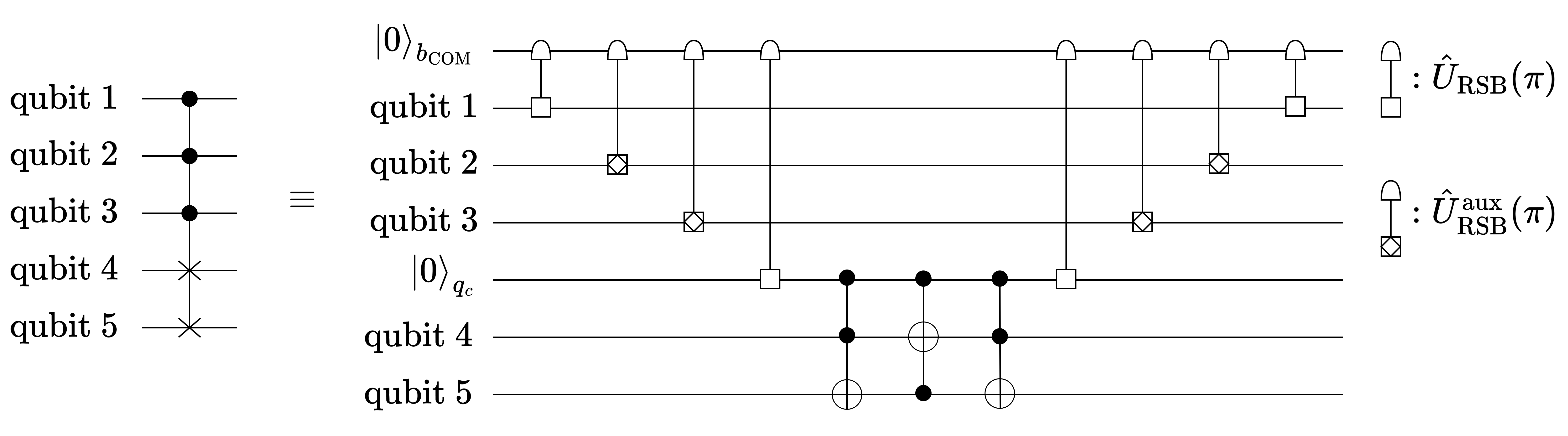}
    \caption{Quantum circuit for implementing three-qubit controlled SWAP gate. Instead of using three \(4\)-CNOT gates, this gate can be decomposed into \(\hat{U}_{\rm{RSB}}(\pi)\), \(\hat{U}_{\rm{RSB}}^{\rm{aux}}(\pi)\) and Toffoli gates.}
    \label{fig: multi cswap}
\end{figure}

One example is the three-qubit controlled SWAP gate (Fig.~\ref{fig: multi cswap}). It is typically decomposed into three \(4\)-CNOT gates. In each \(4\)-CNOT gate, three control qubits are fixed, and the remaining control qubit corresponds to one of the target qubits of the three-qubit controlled SWAP gate. Althernatively, instead of using \(4\)-CNOT gates, three-qubit controlled SWAP gate can alternatively be decomposed into \(\hat{U}_{\rm{RSB}}(\pi)\), \(\hat{U}_{\rm{RSB}}^{\rm{aux}}(\pi)\) and Toffoli gates. Furthermore, if the target qubits are dual-rail qubits, we can use the CSWAP gate described in Sec.~\ref{subsec: cswap} instead of Toffoli gates.

\subsection{Applications of the logical internal qubit--dual-rail qubit hybrid system}
\label{subsec: application}

The logical internal qubit--dual-rail qubit hybrid system scheme can increase the number of available logical qubits compared to conventional trapped-ion quantum computation and offers full connectivity. In this section, we introduce several potential hybrid system applications.

One such application is the simulation of quantum spin systems that are believed to be intractable for classical computation~\cite{fefferman2017exact,park2023hardness}, due to the difficulty of sampling from their output probability distributions—a challenge known as the sampling problem~\cite{Hangleiter}. The proposed spin models require complete balanced bipartite connectivity: spins are divided into two disjoint subsets of equal size, and the Hamiltonian consists of cross-terms between spins from different subsets.

The hybrid system is well suited to such models because its all-to-all connectivity naturally supports complete bipartite connectivity. Furthermore, the hybrid system can simulate a larger spin system than conventional trapped-ion quantum computation, since it provides more available qubits. 

However, measuring all logical qubits in the hybrid system is challenging. In Sec.~\ref{Subsec: prep and detect}, a dual-rail qubit is measured by transferring its information to an internal qubit prepared in \(\ket{\downarrow}\) and subsequently measuring the internal qubit. In the hybrid system, internal qubits also serve as logical qubits, so the logical internal qubits must be measured first. As described in Sec .~\ref {subsec: setting and comparison}, measuring an internal qubit disturbs the quantum information stored in the vibrational modes, affecting the dual-rail qubits. As a result, only the measurement outcomes of logical internal qubits can be reliably obtained.

Nevertheless, we can still leverage the hybrid system by treating dual-rail qubits as auxiliary registers that are not measured. In the simulation of a quantum spin system for the sampling problem, the two disjoint subsets can be assigned to the dual-rail qubits and the logical internal qubits. Notably, the hardness of sampling is preserved even when only the output distribution of the logical internal qubits is considered; thus, it is sufficient to measure only the internal qubits.

\begin{figure}
    \centering
    \includegraphics[width=0.7\linewidth]{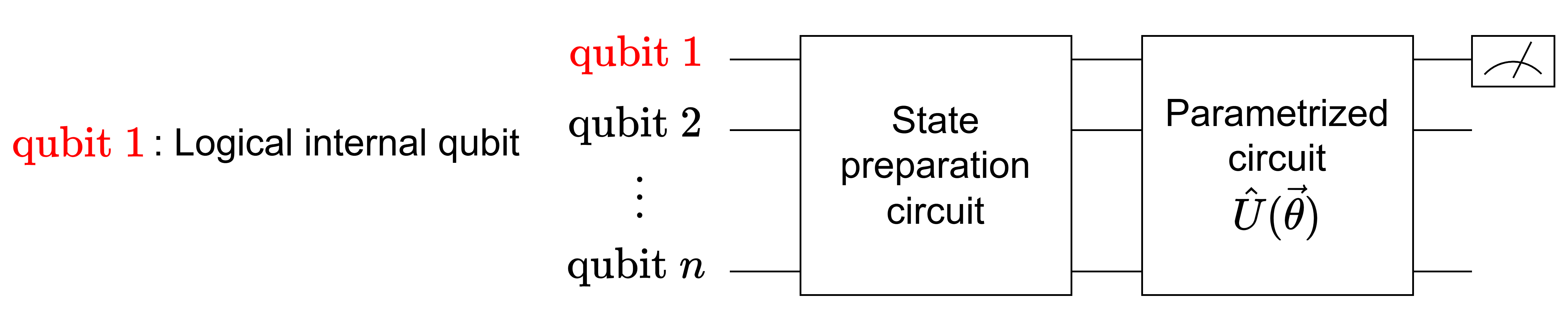}
    \caption{Quantum circuit for a circuit-centric quantum classifier. A logical internal qubit (\(q_1\) in this figure) should be designated as the measurement qubit in the hybrid system. Adapted from Ref.~\cite{schuld2020circuit}.}
    \label{fig: ccqc}
\end{figure}

In addition, many quantum computing applications do not require measuring all qubits. A simple example is the circuit-centric quantum classifier~\cite{schuld2020circuit}, which is a variational quantum algorithm (VQA) in quantum machine learning (QML). This algorithm uses a parameterized quantum circuit to classify classical input data by measuring only one qubit.

Initially, classical input data are encoded into the amplitudes of a quantum state using a state preparation circuit. The parameterized circuit then transforms the encoded information, and the classification is obtained by measuring a single qubit. During the training step, the parameters are optimized to match the expected outputs, and during the prediction step, the quantum circuit provides the classification result. Because the circuit-centric quantum classifier measures only one qubit, it can be implemented within the hybrid system by designating a logical internal qubit as the measurement qubit (Fig.~\ref{fig: ccqc}).

\begin{figure}[htb!]
    \centering
    \includegraphics[width=1.0\linewidth]{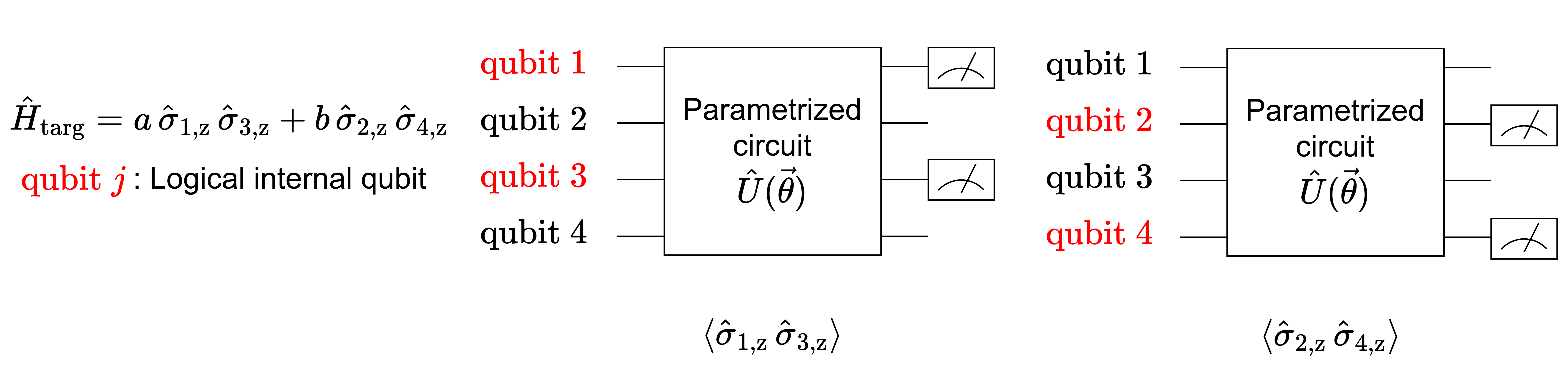}
    \caption{Example of rearrangement of qubit locations for a VQA. Consider a parameterized quantum circuit of four qubits and a target Hamiltonian \(\hat{H}_{\mathrm{targ}}\), which is a linear combination of two Pauli terms: \(\hat{\sigma}_{1,\mathrm{z}}\hat{\sigma}_{3,\mathrm{z}}\) and \(\hat{\sigma}_{2,\mathrm{z}}\hat{\sigma}_{4,\mathrm{z}}\) When a quantum circuit is given, to estimate the expectation value of each Pauli term, the measurement qubits must be designated as logical internal qubits. For \(\hat{\sigma}_{1,\mathrm{z}}\hat{\sigma}_{3,\mathrm{z}}\), \(q_1\) and \(q_3\) are assigned as logical internal qubits; similarly, for \(\hat{\sigma}_{2,\mathrm{z}}\hat{\sigma}_{4,\mathrm{z}}\), \(q_2\) and \(q_4\) are assigned as logical internal qubits.}
    \label{fig: variational}
\end{figure}

Variational quantum eigensolver (VQE)~\cite{peruzzo2014variational} and quantum approximate optimization algorithm (QAOA)~\cite{farhi2014quantum} are VQAs that use parameterized quantum circuits to minimize the expectation value of a target Hamiltonian through parameter optimization. For updating parameters, estimating the expectation value of a target Hamiltonian is required at each step. 

Due to its all-to-all connectivity, the hybrid system is also well-suited for such applications. Generally, a Hamiltonian is a linear combination of Pauli terms, and the expectation value of the Hamiltonian is obtained as a linear combination of the expectation values of its Pauli terms. When a quantum circuit is given, for each Pauli term, we can designate the logical internal qubits as the measurement qubits by rearranging qubit locations, as illustrated in Fig.~\ref{fig: variational}.

Using the rearrangement method, we can exploit the advantages of the hybrid system in studying many-body systems. The hybrid system can treat larger systems compared to the traditional trapped-ion system, and due to its full connectivity, we can simulate many-body systems with various connectivities.

The hybrid system enables simulations of larger many-body systems for studying their properties, particularly in quantum spin dynamics~\cite{lancaster2025simulating}. In quantum spin dynamics, various Hamiltonians, such as the Ising and Heisenberg models, can be represented using local Pauli terms. Since the estimation of the local Pauli terms can be achieved by rearranging qubit locations as in VQAs, the properties of the system with a given state, such as the expectation value of the Hamiltonian or spin-spin correlations, can also be estimated. Consequently, the hybrid system enables simulations of larger systems while retaining the ability to measure physically significant properties.

\section{Conclusion and Discussion}
\label{Sec: conclusion}
In this paper, we have proposed a dual-rail encoding scheme within a trapped-ion phononic network. Using existing operations and interactions, we have described the preparation, measurement, and quantum gate operations on the dual-rail qubits without additional resources. Specifically, we have introduced the ZBS gate and demonstrated how the ZBS gate can implement single-qubit and two-qubit gates. Since adding more ions in a trap can easily increase the number of modes, and all internal qubits and modes can interact directly when the ions are tightly confined, the dual-rail qubit system offers scalability and all-to-all connectivity.

Due to the all-to-all connectivity, a dual-rail qubit system requires only one ancillary internal qubit, while the remaining internal qubits can function as logical qubits. Thus, we have introduced the logical internal qubit--dual-rail qubit hybrid system, which is constructed by incorporating the logical internal qubits into the dual-rail qubit system. We have presented two methods for implementing two-qubit gates between logical internal and dual-rail qubits: CBS gate and RXX gate. Therefore, with existing two-qubit gates between internal qubit states, such as the Mølmer–Sørensen and Cirac-Zoller gates, the all-to-all connectivity property is preserved in the hybrid system. In addition, we have presented how to construct the controlled-SWAP gate using the CBS gates.

We have also proposed an efficient construction of a multi-qubit controlled gate on the hybrid system. Based on the Cirac-Zoller multi-qubit gate protocol, we have presented how to encode the auxiliary state of the logical qubit using a vibrational mode and how to implement the quantum operations on logical qubits for implementing a \(K\)-CNOT gate. We have extended this method from the logical internal qubit system to the hybrid system. This method requires \(\mathcal{O}(K)\) number of operations and \(K\) ancillary modes for \(K\)-CNOT gate. We have noted that since the requirement of ancillary mode reduces the number of dual-rail qubits in the hybrid system, choosing a suitable \(K\) is crucial. A well-chosen \(K\) can ensure that the hybrid system still contains more logical qubits than the traditional quantum computer scheme in a trapped-ion system. Furthermore, we generalized the \(K\)-CNOT gate to an arbitrary multi-qubit controlled gate. By introducing an auxiliary logical internal qubit, we can store the control condition on this logical internal qubit, so a multi-qubit controlled gate can be decomposed into the a total of \(2K+2\) unitaries of the form \(\hat{U}_{\rm{RSB}}(\pi)\) or \(\hat{U}_{\rm{RSB}}^{\rm{aux}}(\pi)\), and a single-qubit controlled gate.

Finally, we have proposed potential applications that utilize the features of the hybrid system. The hybrid system can simulate quantum spin systems with complete bipartite connectivity, which are known to be classically intractable, since it has all-to-all connectivity. Although only logical internal qubits, rather than all logical qubits, can be measured with current techniques, the hybrid system can still be advantageous for certain applications that do not need to measure all qubits. Due to the all-to-all connectivity, the hybrid system can rearrange qubit locations when a quantum circuit and its measurement qubits are given, allowing logical internal qubits to be designated as measurement qubits.

While we have proposed the dual-rail encoding scheme within a phononic network in a trapped-ion system, several considerations remain. First, we have only focused on transverse vibrational modes. Introducing axial modes can also increase the number of available dual-rail qubits; \(N/2\) more dual-rail qubits for \(N\) ions. However, there are several challenges to overcome; the frequencies of axial modes change as the number of ions in a trap increases, which makes the ground state cooling of axial modes difficult, as described in Sec .~\ref {sec: introduction}. Moreover, axial modes have lower frequencies than transverse modes due to the weaker axial trapping potential~\cite{sosnova2021character}.  As a result, controlling axial modes would require different laser frequencies, adding experimental complexity. Investigating how to efficiently integrate axial modes while maintaining high-fidelity gate operations could be an important direction for future research.

Second, off-resonant coupling between internal qubits and vibrational modes can introduce unwanted interactions and degrade gate fidelities, as described in Sec.~\ref{subsec: setting and comparison}. To minimize these effects, careful selection of mode pairs for dual-rail qubits and internal qubits with strong coupling to the chosen modes is necessary. One possible strategy is 'partitioning,' where dual-rail qubits are divided into groups, and each internal qubit is coupled to two groups, which can be strongly coupled. The internal qubit then facilitates quantum gate operations within each group and mediates two-qubit operations both within and across the groups.

Third, a major experimental challenge is that existing measurement techniques disturb the quantum states in the phononic network. This issue prevents the application of total phonon number parity QND measurement circuits during quantum computation. Additionally, this measurement limitation restricts the ability to read out all logical qubits in the hybrid system, as the measurement of internal qubits can collapse the quantum states of dual-rail qubits. Despite these limitations, the dual-rail qubit register remains useful for applications described in Sec.~\ref{subsec: application}.

\begin{acknowledgments}
This work was partly supported by Basic Science Research Program through the National Research Foundation of Korea (NRF), funded by the Ministry of Education, Science and Technology (NRF-2022M3H3A106307411, NRF-2023M3K5A1094813, RS-2024-00413957, and RS-2024-00438415, RS-2025-03532992, RS-2025-07882969). This work was also partly supported by Institute for Information \& communications Technology Promotion (IITP) grant funded by the Korea government (MSIP) (No. 2019-0-00003, Research and Development of Core technologies for Programming, Running, Implementing and Validating of Fault-Tolerant Quantum Computing System). The Ministry of Trade, Industry, and Energy (MOTIE), Korea, also partly supported this research under the Industrial Innovation Infrastructure Development Project (Project No. RS-2024-00466693).  HK is supported by the KIAS Individual Grant No. CG085302 at Korea Institute for Advanced Study. WTC and KK are supported by the Innovation Program for Quantum Science and Technology under Grants No.2021ZD0301602 and the National Natural Science Foundation of China under Grants No.92065205, No.11974200, and No.62335013. JH is supported by the Yonsei University Research Fund of 2025-22-0140.

\end{acknowledgments}

\bibliography{reference}

\appendix

\section{Beamsplitter operation and single-qubit gate implementation on dual-rail encoding}
\label{Appendix_A}

The beamsplitter operator is a bosonic operator that couples two bosonic modes. We recall Eq.~\ref{Eqn: Beamsplitter}:

\begin{align} 
  \hat{B}(\theta,\phi)= 
  \exp\left(i\theta\left(\Cre_{b_1}\Ann_{b_2}e^{i\phi}+\Ann_{b_1}\Cre_{b_2}e^{-i\phi}\right)\right).
  \tag{\ref{Eqn: Beamsplitter}}
\end{align}
For each mode \(b_j \in \{b_1, b_2\}\), the action of the beamsplitter on the creation operator is given by
\begin{gather} \label{Eqn: Beamsplitter On Creation}
\hat{B}(\theta,\phi)\Cre_{b_1}\hat{B}^{\dagger}(\theta,\phi)
= \cos{\theta}\Cre_{b_1} +ie^{-i\phi}\sin{\theta}\Cre_{b_2}, \\
\hat{B}(\theta,\phi)\Cre_{b_2}\hat{B}^{\dagger}(\theta,\phi)
= ie^{i\phi}\sin{\theta}\Cre_{b_1} +\cos{\theta}\Cre_{b_2}.
\end{gather}
Therefore, the action of \(\hat{B}(\theta,\phi)\) on the logical subspace of dual-rail qubit corresponds to the following unitary matrix:
\begin{align}
\hat{B}(\theta,\phi) \Rightarrow
\begin{pmatrix}
    \cos{\theta} & ie^{-i\phi}\sin{\theta} \\
    ie^{i\phi}\sin{\theta} & \cos{\theta}
\end{pmatrix} ,
\end{align}
As shown in Eq.~\ref{eqn: beamsplitter_encoding}, the beamsplitter operation realizes a \(X\cos\phi+Y\sin\phi\)-rotation gate on the dual-rail qubit:
\begin{align}
   \hat{B}(\theta,\phi) \Rightarrow \exp(i\theta(\hat{X}\cos\phi+\hat{Y}\sin\phi)).
   \tag{\ref{eqn: beamsplitter_encoding}}
\end{align}
In particular, \(\hat{B}(\theta/2, \pi)\) and \(\hat{B}(\theta/2,-\pi/2)\) implement \(\hat{R}_X(\theta)\) and \(\hat{R}_Y(\theta)\), respectively. Therefore, an arbitrary single-qubit gate on the dual-rail qubit can be implemented using beamsplitter operations, via the \(X-Y\) decomposition~\cite{barenco1995elementary,nielsen2010quantum}:
\begin{align}
    \hat{U} = e^{i\alpha}\hat{R}_X(\theta_1)\hat{R}_Y(\theta_2)\hat{R}_X(\theta_3), \,\,\, \hat{U} \in \mathrm{SU}(2).
\end{align}

\section{Calculation of TNP phase gate}
\label{Appendix: total number parity phase}
According to Eq.~(\ref{Eqn: Zflip properties}), the operator \(\hat{\mathcal{Z}}\) exchanges the Fock states between two modes with a relative phase that depends on the total phonon number parity:
\begin{align}
\hat{\mathcal{Z}}\Ket{\sigma}\Ket{\underline{n}}\Ket{\underline{m}} = (-\lambda_\sigma i)^{n+m}\Ket{\sigma}\Ket{\underline{m}}\Ket{\underline{n}} = (-\lambda_\sigma)^{(n+m)}i^{n+m}\Ket{\sigma}\Ket{\underline{m}}\Ket{\underline{n}},
    \tag{\ref{Eqn: Zflip properties}}
\end{align}
where \(\sigma \in \{\downarrow,\uparrow\}\). The transformation of the initial state through the TNP phase gate (Fig.~\ref{Fig: total number parity phase gate}) is thus given by: 
\begin{align}
    \Ket{\downarrow}\Ket{\underline{n}}\Ket{\underline{m}} &\eqxrightarrow{\hat{\mathcal{R}}_Y(-\pi/2)} \frac{1}{\sqrt{2}}(\Ket{\downarrow}+\Ket{\uparrow})\Ket{\underline{n}}\Ket{\underline{m}} \\
    & \eqxrightarrow{\hat{\mathcal{Z}}} \frac{i^{(n+m)}}{\sqrt{2}}(\Ket{\downarrow}+(-1)^{(n+m)}\Ket{\uparrow})\Ket{\underline{m}}\Ket{\underline{n}}\\
    &\eqxrightarrow{\hat{\mathcal{R}}_X(\theta)} \exp{\left(i\frac{(-1)^{n+m+1}\theta}{2}\right)}\frac{i^{(n+m)}}{\sqrt{2}}(\Ket{\downarrow}+(-1)^{(n+m)}\Ket{\uparrow})\Ket{\underline{m}}\Ket{\underline{n}} \\
    &\eqxrightarrow{\hat{\mathcal{Z}}^{-1}} \exp{\left(i\frac{(-1)^{n+m+1}\theta}{2}\right)}\frac{1}{\sqrt{2}}(\Ket{\downarrow}+\Ket{\uparrow})\Ket{\underline{n}}\Ket{\underline{m}} \\ 
    &\eqxrightarrow{\hat{\mathcal{R}}_Y(\pi/2)} \exp{\left(i\frac{(-1)^{n+m+1}\theta}{2}\right)}\Ket{\downarrow}\Ket{\underline{n}}\Ket{\underline{m}}.
\end{align}
It shows that the applied phase is \(e^{-i\theta/2}\) if the total number parity \(n+m\) is even, and \(e^{i\theta/2}\) if it is odd. Note that the internal qubit returns to the ground state \(\ket{\downarrow}\), allowing it to be reused for implementing other quantum operations on the modes.

\section{Calculation of the QND total number parity measurement circuit}
\label{Appendix: Bosonic Qubit Error Detection}

\begin{figure}[ht]
\centering
\[ \Qcircuit @C=1em @R=.7em{  
    \lstick{\Ket{\underline{n}}} & \qw      &\multigate{2}{\hat{\mathcal{Z}}}  & \qw      & \multigate{2}{\hat{\mathcal{Z}}} & \qw & \rstick{\Ket{\underline{n}}}\\ 
    \lstick{\Ket{\underline{m}}} & \qw      & \ghost{\hat{\mathcal{Z}}}               & \qw      & \ghost{\hat{\mathcal{Z}}} & \qw & \rstick{\Ket{\underline{m}}}\\
    \lstick{\Ket{\downarrow}} & \gate{\hat{\mathcal{R}}_Y(-\pi/2)} & \ghost{\hat{\mathcal{Z}}}                  & \gate{\hat{\mathcal{R}}_Y(\pi/2)} & \ghost{\hat{\mathcal{Z}}} & \qw & \rstick{\Ket{\sigma_{n \oplus m}}}\\
} \]
\caption{QND measurements for total number parity with the ZBS gate. When the modes are in the Fock states, the internal qubit encodes the total number parity of the two modes without changing their states.}
\label{Fig: QND with ZBS}
\end{figure}
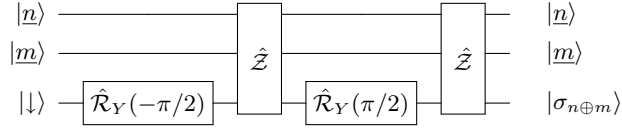

Let an internal qubit be in the ground state \(\Ket{\downarrow}\), while two modes are in Fock states \(\Ket{\underline{n}}\Ket{\underline{m}}\). The sequence of transformations in the QND total number parity measurement circuit (Fig.~\ref{Fig: QND with ZBS}) proceeds as follows:
\begin{align}
    \Ket{\downarrow}\Ket{\underline{n}}\Ket{\underline{m}} &\eqxrightarrow{\quad  \hat{\mathcal{R}}_Y(-\pi/2) \quad \,} \frac{1}{\sqrt{2}} ( \Ket{\downarrow}+\Ket{\uparrow}) \Ket{\underline{n}}\Ket{\underline{m}} \\
    &\eqxrightarrow{\hat{\mathcal{Z}}} \frac{1}{\sqrt{2}}(\Ket{\downarrow}+(-1)^{n+m}\Ket{\uparrow})i^{n+m}\Ket{\underline{m}}\Ket{\underline{n}} \\
    &\eqxrightarrow{\hat{\mathcal{R}}_Y(\pi/2)} i^{(n+m)}\Ket{\sigma_{n\oplus m}}\ket{\underline{m}}\ket{\underline{n}} \\ 
    &\eqxrightarrow{\hat{\mathcal{Z}}}i^{(n+m)}((-\lambda_{\sigma_{n\oplus m}})i)^{(n+m)}\Ket{\sigma_{n\oplus m}}\Ket{\underline{n}}\Ket{\underline{m}}\\
    &\eqxrightarrow{\text{is equivalent to}} \Ket{\sigma_{n\oplus m}}\Ket{\underline{n}}\Ket{\underline{m}},
\end{align}
where \(\oplus\) denotes the addition modulo 2, and
\begin{align}
    \sigma_{n\oplus m} = \begin{cases}
        \downarrow & \text{if } n\oplus m = 0,\\
        \uparrow & \text{if } n\oplus m = 1.
    \end{cases}
\end{align} The final result follows from the identity:
\begin{align}
    i^{(n+m)} \left((-\lambda_{\sigma_{n\oplus m}})i \right)^{(n+m)} = (-1)^{(n+m)}(-\lambda_{\sigma_{n\oplus m}})^{(n+m)} = 1,
\end{align}
Notably, this identity holds for both even and odd values of \((n+m)\), since \(-\lambda_{\sigma_{n\oplus m}}= 1\) when \(n+m\) is even and \(-\lambda_{\sigma_{n\oplus m}}=- 1\) when it is odd.

It can be extended to an arbitrary superposition of product states of two Fock states with the same total number parity:
\begin{align}
 \label{eqn: parity_superpos}
     \Ket{\downarrow}\sum_{n\oplus m = k}\gamma_{n,m}\Ket{\underline{n}}\Ket{\underline{m}} \rightarrow \sum_{n\oplus m = k}\gamma_{n,m}\Ket{\sigma_{n\oplus m}}\Ket{\underline{n}}\Ket{\underline{m}} = \Ket{\sigma_{n\oplus m}}\sum_{n\oplus m = k}\gamma_{n,m}\Ket{\underline{n}}\Ket{\underline{m}}.
\end{align}
Therefore, it can be used to detect errors in a dual-rail qubit caused by phonon number fluctuations, such as single phonon loss or gain. Consider a dual-rail qubit \(D\). In the absence of errors, the total phonon number in \(D\) is \(n+m=1\), so that \( n\oplus m =1\), and the internal qubit state should evolve to \(\ket{\uparrow}\). For example, when a quantum state of \(D\) is given by \(\Ket{\psi}_{D} = \alpha\Ket{0}_{D}+\beta\Ket{1}_{D} \Rightarrow \alpha\Ket{\underline{1}}\Ket{\underline{0}} +\beta\Ket{\underline{0}}\Ket{\underline{1}}\) with \(|\alpha|^2 + |\beta|^2 = 1\), we obtain the following transformation:
\begin{multline}
    \Ket{\downarrow}\Ket{\psi}_{D} \Rightarrow \Ket{\downarrow}(\alpha\ket{\underline{1}}\ket{\underline{0}}+\beta\ket{\underline{0}}\ket{\underline{1}})= \alpha\Ket{\downarrow}\ket{\underline{1}}\ket{\underline{0}}+\beta\Ket{\downarrow}\ket{\underline{0}}\ket{\underline{1}} \\\rightarrow
    \alpha\Ket{\uparrow}\ket{\underline{1}}\ket{\underline{0}}+\beta\Ket{\uparrow}\ket{\underline{0}}\ket{\underline{1}}
    =\Ket{\uparrow}(\alpha\ket{\underline{1}}\ket{\underline{0}}+\beta\ket{\underline{0}}\ket{\underline{1}}) \Rightarrow \Ket{\uparrow}\Ket{\psi}_{D}.
\end{multline}
On the other hand, if a single phonon loss or gain occurs, the total phonon number parity becomes even, causing the internal qubit to transition to \(\ket{\downarrow}\). Consequently, such errors can be detected by measuring the internal qubit, while the quantum state of the dual-rail qubit remains undisturbed in the absence of errors.

\section{Calculation of the RXX gate between internal and dual-rail qubits}
\label{Appendix: rotation XX}

The RXX gate between a logical internal qubit and a dual-rail qubit can be implemented using the ZBS gates. Consider the two-qubit gate realized by the ZBS gate \(\hat{U}_{\mathrm{ZBS}}(\theta,0)\) on a dual-rail qubit and a logical internal qubit from Eq.~\ref{eqn: ZBS_split}:
\begin{align}
    \hat{U}_{\mathrm{ZBS}}(\theta,0) \Rightarrow \ket{0}\bra{0}_q \otimes \hat{R}_X(-2\theta) + \ket{1}\bra{1}_q \otimes \hat{R}_X(2\theta).
\end{align}
The action of the rotation-Y gate \(\hat{R}_Y(\theta)\), which is implemented by \(\hat{\mathcal{R}}_Y(-\theta)\), on the computational basis state of the logical internal qubit \(q\) is given by:
\begin{align}
    \hat{R}_Y(\theta)\ket{x}_q = \cos(\theta/2)\ket{x}_q  +(-1)^x\sin(\theta/2)\ket{x\oplus1}_q,
\end{align}
where \(x \in \{0,1\}\). Thus,
\begin{align}
    \hat{\mathcal{R}}_Y(-\theta)\hat{U}_{\mathrm{ZBS}}(-\pi/4,0) \Rightarrow
    &\,\,\,\cos(\theta/2)\left(\ket{0}\bra{0}_q \otimes \hat{R}_X(\pi/2) + \ket{1}\bra{1}_q \otimes \hat{R}_X(-\pi/2)\right) \\& +\sin(\theta/2)\left(\ket{1}\bra{0}_q \otimes\hat{R}_X(\pi/2)-\ket{0}\bra{1}_q \otimes\hat{R}_X(-\pi/2)\right),
\end{align}
And therefore, the sequence of transformations in the RXX gate circuit (Fig. \ref{Fig: R_xx}) is as follows:
\begin{align}
    \hat{U}_{\mathrm{ZBS}}(\pi/4,0)\hat{\mathcal{R}}_Y(-\theta)\hat{U}_{\mathrm{ZBS}}(-\pi/4,0) \Rightarrow 
    &\,\,\,\cos(\theta/2) \left(\hat{I}_q \otimes \hat{I}_D \right)\\ &+\sin(\theta/2)\left(\ket{1}\bra{0}_q \otimes \hat{R}_X(\pi) - \ket{0}\bra{1}_q \otimes \hat{R}_X(-\pi)\right)\\
    &= \cos(\theta/2)(\hat{I}_q\otimes\hat{I}_D) -i\sin(\theta/2)\hat{X}_q\otimes\hat{X}_D\\
    & = \hat{R}_{XX}(\theta).
\end{align}
This result represents the RXX gate between a logical internal qubit and a dual-rail qubit.

\end{document}